# Improved Method for Searching of Interleavers Using Garello's Method

Lucian Trifina and Daniela Tarniceriu

***Abstract*—In this paper an improved method for searching good interleavers from a certain set is proposed. The first few terms, corresponding to maximum distance of approximately 40 of the distance spectra, for turbo codes using these interleavers are determined by means of Garello's method. The method is applied to find quadratic permutation polynomials (QPP) based interleavers. Compared to previous methods for founding QPP based interleavers, the search complexity is reduced, allowing to find interleavers of higher length.**

**This method has been applied for QPP interleavers with lengths from the LTE (Long Term Evolution) standard up to 1504. The analyzed classes are those with the largest spread QPP (LS-QPP), with the D parameter equal to that of LTE interleaver ($D_{LTE}$-QPP), and the class consisting of all QPP interleavers for lengths up to 1008.**

**The distance spectrum optimization is made for all classes. For the class of LS-QPP interleavers of small lengths, the search led to superior or at least equal performances with those of the LTE standard. For larger lengths the search in the class of DLTE-QPP interleavers is preferred. The interleavers from the entire class of QPPs lead, in general, to weaker FER (Frame Error Rate) performance.**

# I. INTRODUCTION

Finding good interleavers for turbo codes is a crucial requirement to improve their performance. Optimizing the distance spectrum is a widely used method to accomplish this.

To determine exactly the distance spectrum, the well known Garello's method is used [1]. However, for large interleaver lengths and distances higher than about 40, it is very computational demanding.

Even for smaller distances the search of interleavers by optimizing the distance spectrum can pose problems when the number of interleavers in a certain class is too large. This has motivated the reduction of search complexity by the method presented in this paper.

In the following, we present QPP (Quadratic Permutation Polynomials) based interleavers which have been proposed by Sun and Takeshita [2] and which are used in LTE (Long Term Evolution) standard [3], due to their generation simplicity and the ability to result in very good performances of bit and/or frame error rates (BER / FER).

A QPP interleaver of length $L$ is described by:

$$\pi(x) = \left(q_0 + q_1 x + q_1 x^2\right) \bmod L, x = 0,1,\ldots,L-1, \tag{1}$$

where $q_1$ and $q_2$ are chosen so that the quadratic polynomial in (1) is a permutation polynomial and $q_0$ only determines a shift of the permutation elements. In the LTE standard only QPPs with $q_0 = 0$ are considered.

In [4] it was proved that there are lengths in the LTE standard for which the QPP based interleavers can be improved.

In [5] Takeshita gave examples of good interleavers with the largest $D$ parameter or spreading factor. It is defined as:

$$D = \min_{\substack{i \neq j \\ i,j \in \mathbb{Z}_L}} \left\{ \delta_L \left( p_i, p_j \right) \right\} \tag{2}$$

where $\mathbb{Z}_L = \{0,1,\ldots,L-1\}$ and $\delta_L(p_i, p_j)$ is the Lee metric between points $p_i = (i, \pi(i))$ and $p_j = (j, \pi(j))$:

$$\delta_L(p_i, p_j) = |i-j|_L + |\pi(i) - \pi(j)|_L, \tag{3}$$

where

$$|i-j|_L = \min\{(i-j)(\text{mod}\, L), (j-i)(\text{mod}\, L)\}. \tag{4}$$

The randomizing analysis of these interleavers was made considering the nonlinearity degree [5]. It is demonstrated that the nonlinearity degree of a QPP interleaver is given by [5]:

$$\zeta = L / \text{gcd}(2q_2, L), \tag{5}$$

where “gcd” means “greatest common divisor”.

The refined nonlinearity degree $\zeta'$ is given by the number of distinct elements of the set $\{q_2 x^2 \text{ (mod } L),\ x=0,1, ..., \zeta-1\}$ and is a more appropriate measure for the randomness of QPP interleavers.

For short lengths the method used in [4] consists in searching among QPP interleavers with the largest $D$ parameter those leading to the best distance spectrum (considering the first few terms). These interleavers lead to lowest truncated upper bounds (TUB) of FER. The values of TUB for BER and FER for independent fading channel (as wireless channel) are given by [6]:

$$TUB(BER) = 0.5 \cdot \sum_{i=1}^{M} \frac{w_i}{L} \cdot \left( \frac{1}{1 + R_c \cdot SNR} \right)^{d_i}, \tag{6}$$

$$TUB(FER) = 0.5 \cdot \sum_{i=1}^{M} N_i \cdot \left( \frac{1}{1 + R_c \cdot SNR} \right)^{d_i}, \tag{7}$$

where $M$ is the number of terms in the distance spectrum taken into account, $d_i$ is the $i^{th}$ distance in the spectrum, $w_i$ is the total information weight corresponding to distance $d_i$, $N_i$ is the number of code words with distance $d_i$, $R_c$ is the coding rate and $SNR$ is the signal to noise ratio. The trellis termination used in the LTE standard is post interleaver, transmitting the termination bits of the second trellis. Since the turbo code in the LTE standard uses a component code with memory 3, the coding rate is calculated by:

$$R_c = \frac{L}{3 \cdot L + 12} \tag{8}$$

However, for medium and large lengths the complexity of this method has to be reduced in order to find better QPPs in a reasonable time interval. This paper proposes an improvement of the method in [4], by reducing its complexity, allowing its application also for longer lengths.

The paper is structured as follows. Section II presents the improved method for searching interleavers from a certain class and numerical results regarding the QPP coefficients are given by searching in three QPP classes. These numerical results consist in: QPP coefficients, TUB(BER) and TUB(FER) values, the *D* parameter, and the number of polynomials resulted for a certain length from that class. Section III presents simulation results for several cases, and section IV concludes the paper.

## II. Improved Method and Numerical Results

The method used in [4] to find QPPs with improved performances involved calculation of distance spectra using Garello's method [1]. This was applied for each QPP (as in the first method in [7]) or for a group of QPPs leading to identical permutations, with the largest spreading factor (*D*) (according to the condition provided in [8] and independently in [4]).

A reduction of the number of QPPs (or, in general, of the permutation polynomials) for which the distance spectrum is calculated has been given in [9]. One can note that for classical symmetric turbo codes (identical recursive convolutional codes parallel concatenated) the inverse interleaver results in the same distance spectrum. Thus, if a QPP admits an inverse QPP, different from the initial one or from one equivalent to it, then the distance spectrum was calculated only once for each group of such four QPPs.

In this paper, we propose an improvement of this method of interleaver search from a certain set (particularly, in a QPP set), speculating Garello's algorithm, to determine the distance spectrum. As proposed in [1] or implemented in C language in [10], the algorithm involves the update of the first few terms of the distance spectrum by following the trellis corresponding to the component convolutional code from its end towards its beginning and achieving constrained subcodes.

In the following we present the search algorithm particularized for QPP based interleavers, although it can be applied to any other class of interleavers.

After finding all QPPs from a certain set, we determine the distance spectrum using the complete Garello's algorithm for the first group of QPPs, equivalent in terms of distance spectrum. The corresponding TUB(FER) value is denoted by $FER_{min}$. For the other groups of QPPs, equivalent in terms of distance spectrum, we calculate the value of TUB(FER) at each update of the distance spectrum in Garello's algorithm.

If the first $M$ distances $d_1, d_2, \ldots, d_M$ have to be calculated, then, to avoid the situation when a spectrum leading to a lower TUB(FER) exists at a step closer to the beginning of the trellis, we check at each step if the improving of distance spectrum with $M$ terms leads to a lower TUB(FER). In the first case the update of the spectrum is made as

$$d_{1_new} = d_1 - 1, d_{2_new} = d_1, \ldots, d_{M_new} = d_{M-1} \tag{9}$$

The multiplicity for the first distance is set to the most favorable case, i.e. 1, and the other $M-1$ multiplicities are updated similar to the distances, that is

$$N_{1_new} = 1, N_{2_new} = N_1, \ldots, N_{M_new} = N_{M-1} \tag{10}$$

For a TUB(FER) larger than $FER_{min}$, the spectrum update continues until

$$d_{1_new} = d_1 - M, d_{2_new} = d_1 - (M-1), \ldots, d_{M_new} = d_1 - 1 \tag{11}$$

with multiplicities

$$N_{1_new} = 1, N_{2_new} = 1, \ldots, N_{M_new} = 1 \tag{12}$$

This is the last from the most optimistic cases of spectrum update.

If for one of these spectrum updates a TUB(FER) lower than $FER_{min}$ results, we continue with spectrum determination at the next step in the trellis. In this way we can get the complete distance spectrum and the corresponding value of TUB(FER) will be assigned to $FER_{min}$. Otherwise, no other spectrum update will lead to a lower TUB(FER) and we quit calculating the distance spectrum and move to the next group of equivalent QPPs. The method can be easily implemented by a simple change of the function "*gamma*" in [10].

The complexity reduces in this way, especially if the QPPs from the set in which we search are in descending order of TUB(FER) performances and if the error patterns that lead to the smallest distances are located towards the end of the trellis. The search is performed by computing the distance spectra.

Below we present the search results for three classes of QPPs. These classes are those with the largest spread QPP (LS-QPP), with the $D$ parameter equal to that of LTE interleaver ($D_{LTE}$-QPP), and the class consisting of all QPP interleavers for lengths up to 1008.

Table I gives the QPPs in the LTE standard and those found out by optimizing the distance spectrum for Rayleigh fading channel among the largest spread QPPs (LS-QPPs) set, using the method previously described. The considered lengths are from 40 to 1504. The results for improved interleavers for lengths up to 512 are given in [4]).

For the specified SNR values and the considered number of distances, the values $10^7$×TUB(BER) and $10^5$×TUB(FER), the value of the $D$ parameter, the minimum distances ($d_{min}$) and their multiplicities ($N_1$ and $w_1$) for each QPP based interleaver are also given. The penultimate column gives the number of polynomials which lead to the highest value of the $D$ parameter and minimum TUB(FER) for that length. The table only presents polynomials with the lowest $q_1$ and then with the lowest $q_2$. In the last column the ratio between the TUB(FER) for the LTE interleaver (denoted FER_LTE) and that found by the proposed method ($FER_{min}$) is given.

Since the length of QPP based interleaver is increased (having influence on the computing time in Garello's method), the number of distances was considered at least 3 until the length of 816. For greater lengths only the first term in the distance spectrum was taken into account.

From Table I we can see that in many cases the resulting TUB(FER) are significantly lower than those for LTE-QPPs (lengths marked in gray). However, this does not mean better FER performance for longer lengths, where another approach should be considered. For large lengths the maximization of the *D* parameter is not important, but we have to take into account the multiplicities in the distance spectrum [5].

As in [4], when the $FER_{min}$ value results greater than FER_LTE, the search is done over all QPPs having the parameter *D* greater or equal to that for LTE-QPP. The results are given in Table II.

In Table III the searching times for the previous method and the new improved one are given for several lengths. We note that in the proposed method the searching time is significantly reduced, meaning reduced search complexity. However, it should be noted that the search time for a certain length depends on the number of QPPs with largest *D* parameter. For longer lengths and greater number of QPPs with the largest *D* parameter the time required by the previous method is just too large, so the new method is a salutary solution. Similar results in terms of searching time can also be given for other classes of interleavers than LS-QPP.

For longer lengths the simulation results did not show any improvement for the LS-QPP class. We have considered the QPP class with the *D* parameter as in the LTE standard ($D_{LTE}$-QPP) and the highest $\zeta'$. The condition to have the highest $\zeta'$ ensures a lower multiplicity and reduces the number of QPPs from the class. However, it can also affect the performance for smaller lengths. The results are given in Table IV. For comparison with the results in [4], the search results for lengths from 40 to 512 are also given. This table presents the same quantities as Table I, plus the value of $\zeta'$, for both interleavers at each length.

In further analysis we compute the ratio between the TUB(FER) values of the first interleaver and the second one. The second interleaver is considered to be better than the first one if the ratio is higher than 1, weaker if the ratio is less than 1, and equally good when the ratio is equal to 1. When the ratio of TUB (FER) values is greater than 2, it means that there is a significant improvement (proven in [4] by simulation for length 40).

From Table IV, among the first 60 lengths (from 40 up to 512), we see that for 9 lengths (i.e. 40, 128, 240, 328, 384, 408, 424, 456 and 488) significantly improved interleavers were obtained compared to those in the LTE standard and equally good as the LS-QPPs-TUB (FER)min, given in [3]. We mention that in Table I of [3] LS-QPPs-TUB(FER) interleavers were given for 19 such lengths (length 328 was omitted in [3]).

For 3 lengths (i.e. 400, 464 and 496) interleavers better than those in the LTE standard and better than the LS-QPPs-TUB (FER)min were obtained. For length 464 the improvement is not significant, as we can see in the table, the ratio of the TUB(FER) values being less than 2.

For 2 lengths (280 and 440) interleavers as good as those in the LTE standard and better than those LS-QPPs-TUB(FER)min were obtained.

For 9 lengths (i.e. 184, 216, 256, 288, 320, 352, 416, 448 and 480) interleavers better than those of the LTE standard were obtained, but weaker than those LS-QPPs-TUB(FER)min. Only for 3 of these lengths (184, 256 and 416) the ratio of the TUB(FER) values is greater than 2.

For a single length (i.e. 504) an interleaver as good as one in the LTE standard, but weaker than that LS-QPPs-TUB (FER)min was obtained.

For a single length (i.e. 72) an interleaver weaker than that in the LTE standard was obtained, but better than that LS-QPPs-TUB(FER)min.

For a single length (i.e. 168) an interleaver weaker than that in the LTE standard and as weak as that LS-QPPs-TUB(FER)min was obtained. We mention that for this length the LTE-QPP interleaver is reducible to a linear one, and the search was made only among QPP interleavers irreducible to linear permutation polynomial (LPP) based interleavers.

For a single length (i.e. 144) an interleaver weaker than that in the LTE standard and that LS-QPPs-

TUB(FER)min was obtained.

Finally, Table V gives the QPP interleavers with lengths from the LTE standard from 40 up to 1008, leading to minimum TUB(FER) for a number of distances in distance spectrum, among all QPP interleavers of a given length (denoted by QPP-TUB(FER)min interleavers).

Figure 1 shows the distances obtained with these QPP-TUB(FER)min interleavers, along with those obtained with LTE-QPP and the $D_{LTE}$-QPP-TUB(FER)min, when used in the turbo code with LTE component convolutional code. We note an upper bound of 35 for the minimum distance of QPP-TUB(FER)min interleavers for lengths up to 1008. In [11] it is shown that the largest minimum distance for QPP interleavers with lengths up to 1008 and dual termination is 46, which indicates that the LTE termination (i.e. post-interleaver) decreases the minimum distance of turbo codes.

However, the LTE termination is simpler and more used in practice. Of the first 60 lengths (from 40 to 512) the same *D* parameter has resulted as for LS-QPP only for six lengths (i.e. lengths of 128, 256, 328, 352, 456, 480). It should be noted that although TUB(FER) is minimum, these interleavers do not necessarily lead to the best BER/FER performance. A cause is the *D* parameter different from that corresponding to LTE-QPP, which is either too small or too large. This influences the terms in the distance spectrum superior to those considered.

The minimum distance obtained with $D_{LTE}$-QPP- TUB(FER)min interleavers is greater than or equal to that obtained with LTE-QPP ones (excepting the length of 1008, when the distance is slightly less, 29 comparing to 30, because of lower multiplicity). The above minimum distance is less than or equal to that resulted with QPP-TUB(FER)min interleavers. For lengths greater than or equal to 488, the minimum distance resulted with $D_{LTE}$-QPP-TUB(FER)min interleavers is lower than that resulted with QPP-TUB(FER)min interleavers, excepting three lengths. These are 560 and 720, when the resulting distances are equal, and 672, when the distance is greater with 1. However, as we shall see in the next section, the FER performances obtained by simulation with $D_{LTE}$-QPP-TUB(FER)min interleavers is superior to those of QPP-TUB(FER)min ones, showing the importance of the D parameter, besides the first terms of distance spectrum.

For a better view, in Figure 2 the D parameter was drawn for LTE-QPP, QPP-TUB(FER)min and LS-QPP interleavers. We note that for lengths until approximately 800 (excepting 17 from the 78 lengths, namely 184, 280, 368, 400, 440, 464, 496, 528, 560, 592, 624, 656, 688, 704, 752, 768 and 784) the *D* parameter of LTE-QPP interleavers is the largest possible (i.e. that of the LS-QPP interleavers). For higher lengths the D parameter is lower.

As stated in [5], this can be explained by the fact that in designing good LTE interleavers we take into account that for smaller lengths a larger *D* parameter is more important and for larger lengths the maximization of the metric $\Omega'$ is more important. The maximization implies a compromise between the *D* parameter and the refined nonlinearity degree $\zeta'$, which implies a *D* parameter smaller than the largest possible.

For the remaining 29 lengths (up to 1504) only four of these LTE interleavers have the parameter *D* equal to that of LS-QPP interleavers, namely, the lengths of 216, 1280, 1408 and 1472. Most of times the LS-QPP-TUB(FER)min interleavers have the *D* parameter significantly greater, which affects the performance of turbo codes with these interleavers.

## III. Simulation Results

In this section simulation results are given in some representative cases for four interleaver lengths. We simulated LTE and TUB(FER)min interleavers from the following classes:

- with the largest *D* parameter (LS-QPP);
- with the *D* parameter equal to that of the LTE interleaver ($D_{LTE}$-QPP);
- all QPP interleavers for a certain length.

All the considered QPP interleavers are irreducible to LPP interleavers.

The channel is modeled with independent Rayleigh fading. The decoding algorithm is Log-MAP, with iteration stop criterion based on LLR magnitude (the iterations are stopped when the magnitudes of all the LLR values are greater than a threshold equal to 10). At least 100 erroneous frames are simulated, excepting for some high SNR values, when the number of erroneous frames was still greater than 50. The figures containing the simulated FER curves also show the *D* parameters of each interleaver.

The lengths of interleavers we have considered are shown below.

A. The length of the simulated interleavers is 448. Their performances are shown in Figure 3.

We note that the best performance is achieved by the LS-QPP interleaver with minimum TUB(FER), followed by the interleaver with the best distance spectrum among all interleavers, and then by the LTE interleaver. Although the minimum distance of the second interleaver is higher than that of the LS-QPP with minimum TUB(FER) (28 vs. 25), its FER performance is weaker. An explanation is given by the interpretation of TUB(FER) curves for 3 and 9 terms of distance spectrum, given in Figure 4. We note that for LS-QPP-TUB(FER)min and LTE interleavers (which, for this length also belong to the LS-QPP class) the TUB(FER) curve is almost the same, while for the QPP-TUB(FER)min interleaver this quantity is significantly smaller for 9 terms comparing to that for 3 terms. This happens due to large multiplicities for large distances from the minimum distance. This fact is influenced by the larger *D* parameter for the LS-QPP interleaver compared to that of the QPP-TUB(FER)min (28 vs. 14). The fact that the LTE interleaver has the same *D* parameter as the LS-QPP-TUB(FER)min, but significantly weaker performance, justifies the search among interleavers with the same *D* parameter, by improving the distance spectrum with a reasonable number of terms.

The $D_{LTE}$-QPP-TUB(FER)min interleaver was not longer simulated for this length, because it has the same *D* and ζ' parameters as the LTE interleaver and leads to the same minimum distance and the same multiplicities.

B. The length of the simulated interleavers is 624. Their performances are shown in Figure 5.

We notice that the best performance is obtained by the LTE interleavers with the best distance spectrum for the same parameter $D$ = 22 ($D_{LTE}$-QPP- TUB(FER)min). The interleaver with the best distance spectrum among all interleavers has the parameter $D$ = 18 and the weakest performance, while the LS-QPP interleaver with the best distance spectrum with parameter $D$ = 26 (the largest) has slightly better performance. We mention that the interleaver with the best distance spectrum among all interleavers leads to the minimum distance 31, with multiplicity 1. This minimum distance is the largest among the simulated interleavers of length 624. This shows that only the minimum distance or even the distance spectrum with a small number of terms is not a good indicator for turbo code performance. Conveniently choosing the $D$ parameter significantly improves the performance.

C. The length of the simulated interleavers is 1008. Their performances are shown in Figure 6.

Although the TUB(FER) for the $D_{LTE}$-QPP interleaver with minimum TUB(FER) is significantly lower than that for the LTE interleaver (see Table IV) for one term in distance spectrum, in simulation the performance difference is small. This is because the second term in the distance spectrum for the searched interleaver with the same $D$ parameter as the LTE interleaver is equal to the first term in the distance spectrum for the the second interleaver (i.e. 30/322/644), thus reducing the performance of the first one.

The performance of the interleaver with the best distance spectrum (one term) from all the QPP interleavers of length 1008 is significantly weaker. This time its $D$ parameter is larger, that is $D$ = 36. The performance of LS-QPP interleaver with the highest parameter $D$ ($D$ = 42) is similar to that of the interleaver with parameter $D$ = 36, slightly lower at high SNR.

D. The length of the simulated interleavers is 832. Their performances are shown in Figure 7.

Note that the $D_{LTE}$-QPP-TUB(FER)min interleaver leads to superior performance than the LTE one at high SNR (lower error floor), because of the same parameters D and $\zeta'$ and larger minimum distance (27 vs 23).

## IV. Conclusions

An improved method to search interleavers from a certain set was proposed using Garello's method to determine the distance spectrum. Its complexity is reduced, allowing to find interleavers for lengths up to about 1500 in a reasonable time period for sets containing up to few hundreds interleavers.

The search method was applied to QPP interleavers from the classes LS-QPP, QPP interleavers with the *D* parameter equal to that in LTE standard and highest refined nonlinearity degree and, finally, the whole class of QPP interleavers. Numerical results are given for the first two classes up to the length 1504, and for the last one up to the length 1008. It was found that the first 78 lengths (up to the length 800), excepting 17 lengths, the LTE interleaver is part of the class LS-QPP, making the QPP search more suitable by optimizing distance spectrum in this class.

With few exceptions, the LS-QPP interleavers with larger lengths lead to weaker BER/FER performances, so that the search in the QPP class with the *D* parameter equal to that of LTE interleaver is more appropriate.

The search in the class of all QPP interleavers, by optimizing the distance spectrum with a reasonable number of terms, do not lead most of the times to interleavers superior in terms of FER performance, because their *D* parameter is too small. The simulation results in Section III, for some representative cases, confirm what was previously stated.

Finally, it should be noted that the improved search method of interleavers from a certain class, optimized by distance spectrum can also be applied for known methods for estimating the minimum distance, when Garello’s method is very computational demanding. These methods are double or triple impulse methods (DIM, TIM) [12], and full range double or triple impulse methods (DIMK, TIMK, TIMKK) [13].

## References

[1] Garello R, Pierleoni P, Benedetto S. Computing the Free Distance of Turbo Codes and Serially Concatenated Codes with Interleavers: Algorithms and Applications, *IEEE Journal on Selected Areas in Communications* **2001; 19(5)**, pp. 800-812, DOI: 10.1109/49.924864.

[2] Sun J, Takeshita OY. Interleavers for Turbo Codes Using Permutation Polynomial Over Integers Rings, *IEEE Transactions on Information Theory* **2005; 51(1)**, pp. 101-119, DOI: 10.1109/TIT.2004.839478.

[3] 3GPP TS36.212, v8.3.0 (2008-05), Multiplexing and channel coding, 2008; 12-14.

[4] Trifina L, Tarniceriu D, Munteanu V. Improved QPP Interleavers for LTE Standard, *IEEE International Symposium on Signals, Circuits and Systems ISSCS 2011* **2011;** pp. 403-406, DOI: 10.1109/ISSCS.2011.5978745.

[5] Takeshita OY. Permutation Polynomial Interleavers: An Algebraic-Geometric Perspective, *IEEE Transactions on Information Theory* **2007; 53(6),** pp. 2116-2132, DOI: 10.1109/TIT.2007.896870.

[6] Yuan J, Feng W, Vucetic B. Performance of Parallel and Serial Concatenated Codes on Fading Channels, *IEEE Transactions on Communication*s **2002; 50(10)**, pp. 1600-1608, DOI: 10.1109/TCOMM.2002.803971.

[7] Tarniceriu D, Trifina L, Munteanu V. About Minimum Distance for QPP Interleavers, *Annals of Telecommunications* **2009; 64(11-12)**, pp. 745-751, DOI: 10.1007/s12243-009-0120-3.

[8] Zhao H, Fan P, Tarokh V. On the Equivalence of Interleavers for Turbo Codes Using Quadratic Permutation Polynomials over Integer Rings, *IEEE Communications Letters* **2010; 14(3)**, pp. 236-238, DOI: 10.1109/LCOMM.2010.03.091695.

[9] Trifina L, Tarniceriu D. Analysis of Cubic Permutation Polynomials for Turbo Codes, *Wireless Personal Communications*, accepted for publication. Available: http://arxiv.org/vc/arxiv/papers/1106/1106.1953v4.pdf

[10] http://www.tlc.polito.it/garello/turbodistance/turbodistance.html

[11] Rosnes E. On the Minimum Distance of Turbo Codes With Quadratic Permutation Polynomial Interleavers, *IEEE Transactions on Information Theory*, accepted for publication, see the next link for the first version http://arxiv.org/PS_cache/arxiv/pdf/1102/1102.5275v1.pdf

[12] Crozier S, Guinand P, Hunt A. Estimating the Minimum Distance of Turbo-Codes Using Double and Triple Impulse Methods, *IEEE Communications Letters* **2005; 9(7)**, pp. 631-633, DOI: 10.1109/LCOMM.2005.1461687.

[13] Crozier S, Guinand P, Hunt A. Estimating the Minimum Distance of Large-Block Turbo Codes using Iterative Multiple-Impulse Methods, *European Transactions on Telecommunications* **2007; 18(5)**, pp. 437-444, DOI: 10.1002/ett.1185.

TABLE I
LTE-QPP AND LS-QPP WITH MINIMUM TUB(FER) INTERLEAVERS

| | | | LTE- QPP Interleavers | | | | | LS-QPP- TUB(FER)min Interleavers | | | | | | |
|---|---|---|---|---|---|---|---|---|---|---|---|---|---|---|
| L | SNR [dB] | num dist | $\pi(x)$ | D | $d_{min}/N_1/w_1$ | TUB (BER) $*10^7$ | TUB (FER) $*10^5$ | $\pi(x)$ | D | $d_{min}/N_1/w_1$ | TUB (BER) $*10^7$ | TUB (FER) $*10^5$ | No. pol. | $FER_LTE/FER_{min}$ |
| 40 | 7.5 | 9 | $3x+ 10x^2$ | 4 | 11/1/3 | 10.559 | 1.6211 | $13x+ 30x^2$ | 4 | 12/1/2 | 4.0451 | 0.6539 | 4 | 2.48 |
| 48 | 7.5 | 9 | $7x+ 12x^2$ | 6 | 13/1/3 | 1.1890 | 0.1838 | $7x+ 36x^2$ | 6 | 15/2/6 | 0.7589 | 0.1150 | 2 | 1.60 |
| 56 | 7.5 | 9 | $19x+ 42x^2$ | 6 | 13/1/1 | 3.3169 | 0.9523 | $3x+ 42x^2$ | 6 | 13/1/1 | 3.3169 | 0.9523 | 4 | 1.00 |
| 64 | 7.5 | 9 | $7x+ 16x^2$ | 8 | 12/1/2 | 1.0947 | 0.3523 | $9x+ 48x^2$ | 8 | 12/1/2 | 1.1002 | 0.3456 | 4 | 1.02 |
| 72 | 7.5 | 9 | $7x+ 18x^2$ | 6 | 15/1/1 | 0.0521 | 0.0236 | $5x+ 60x^2$ | 8 | 12/2/4 | 1.6174 | 0.5677 | 4 | 0.04 |
| 80 | 6.5 | 9 | $11x+ 20x^2$ | 10 | 19/3/5 | 0.1369 | 0.0344 | $11x+ 20x^2$ | 10 | 19/3/5 | 0.1369 | 0.0344 | 4 | 1.00 |
| 88 | 6.5 | 9 | $5x+ 22x^2$ | 8 | 15/1/1 | 0.5231 | 0.2584 | $5x+ 22x^2$ | 8 | 15/1/1 | 0.5231 | 0.2584 | 4 | 1.00 |
| 96 | 6.5 | 9 | $11x+ 24x^2$ | 12 | 16/2/4 | 0.3015 | 0.1325 | $13x+ 72x^2$ | 12 | 16/1/2 | 0.2173 | 0.0942 | 4 | 1.41 |
| 104 | 6 | 9 | $7x+ 26x^2$ | 8 | 16/1/2 | 0.3928 | 0.2403 | $37x+ 26x^2$ | 8 | 16/1/2 | 0.4179 | 0.2028 | 4 | 1.18 |
| 112 | 6 | 9 | $41x+ 84x^2$ | 14 | 17/2/2 | 0.3613 | 0.2200 | $41x+ 28x^2$ | 14 | 17/2/2 | 0.3613 | 0.2200 | 4 | 1.00 |
| 120 | 6 | 7 | $103x+ 90x^2$ | 10 | 16/1/2 | 0.3376 | 0.2351 | $17x+ 90x^2$ | 10 | 16/1/2 | 0.3045 | 0.1609 | 4 | 1.46 |
| 128 | 5.5 | 7 | $15x+ 32x^2$ | 16 | 16/12 | 1.2349 | 0.6560 | $17x+ 32x^2$ | 16 | 18/1/2 | 0.2189 | 0.1446 | 4 | 4.54 |
| 136 | 5.5 | 7 | $9x+ 34x^2$ | 10 | 16/3/4 | 1.6234 | 1.0085 | $19x+ 102x^2$ | 10 | 16/1/2 | 0.9306 | 0.5515 | 4 | 1.83 |
| 144 | 5 | 7 | $17x+ 108x^2$ | 16 | 20/2/4 | 0.4829 | 0.2873 | $19x+ 36x^2$ | 16 | 19/1/1 | 0.2131 | 0.1431 | 4 | 2.01 |
| 152 | 5 | 7 | $9x+ 38x^2$ | 12 | 15/1/1 | 2.9059 | 2.6074 | $59x+ 38x^2$ | 12 | 16/1/2 | 2.2269 | 1.4680 | 4 | 1.78 |
| 160 | 5 | 7 | $21x+ 120x^2$ | 16 | 19/1/1 | 0.2742 | 0.2535 | $19x+ 120x^2$ | 16 | 19/1//1 | 0.2383 | 0.2274 | 4 | 1.11 |
| 168 | 5 | 7 | $101x+ 84x^2$ | 12 | 17/1/1 | 0.2745 | 0.3644 | $61x+ 126x^2$ | 12 | 21/1/1 | 1.4998 | 1.2596 | 4 | 0.29 |
| 176 | 5 | 7 | $21x+ 44x^2$ | 16 | 20/2/4 | 0.2018 | 0.1691 | $21x+ 44x^2$ | 16 | 20/2/4 | 0.2018 | 0.1691 | 2 | 1.00 |
| 184 | 5 | 7 | $57x+ 46x^2$ | 12 | 16/1/2 | 1.0900 | 0.9958 | $25x+ 46x^2$ | 14 | 20/2/4 | 0.1083 | 0.0959 | 4 | 10.38 |
| 192 | 4.5 | 7 | $23x+ 48x^2$ | 16 | 22/1/2 | 0.2096 | 0.1839 | $23x+ 144x^2$ | 16 | 22/1/2 | 0.1878 | 0.1735 | 4 | 1.06 |
| 200 | 4.5 | 7 | $13x+ 50x^2$ | 14 | 20/1/2 | 0.5008 | 0.4360 | $13x+ 150x^2$ | 14 | 20/1/2 | 0.4798 | 0.3839 | 4 | 1.14 |
| 208 | 4.5 | 7 | $27x+ 52x^2$ | 16 | 23/1/1 | 0.1889 | 0.1546 | $25x+ 52x^2$ | 16 | 23/1/1 | 0.1889 | 0.1546 | 4 | 1.00 |
| 216 | 4.5. | 7 | $11x+ 36x^2$ | 18 | 22/1/2 | 0.1029 | 0.1210 | $23x+ 144x^2$ | 18 | 22/1/2 | 0.0925 | 0.1038 | 4 | 1.17 |
| 224 | 4.5 | 7 | $27x+ 56x^2$ | 16 | 22/99 /198 | 2.5241 | 2.8293 | $27x+ 168x^2$ | 16 | 22/98 /196 | 2.5039 | 2.8129 | 4 | 1.01 |
| 232 | 4.5 | 7 | $85x+ 58x^2$ | 16 | 23/1/1 | 0.0226 | 0.0297 | $15x+ 174x^2$ | 16 | 23/1/1 | 0.0252 | 0.0295 | 4 | 1.01 |
| 240 | 4.5 | 7 | $29x+ 60x^2$ | 16 | 24/2/4 | 0.5208 | 0.4262 | $89x+ 60x^2$ | 16 | 24/1/2 | 0.0897 | 0.0807 | 4 | 5.28 |
| 248 | 4.5 | 7 | $33x+ 62x^2$ | 18 | 23/2/4 | 0.0451 | 0.0629 | $33x+ 186x^2$ | 18 | 23/1/1 | 0.0336 | 0.0406 | 4 | 1.55 |
| 256 | 4.5 | 7 | $15x+ 32x^2$ | 16 | 16/1/2 | 1.3491 | 1.7748 | $31x+ 192x^2$ | 16 | 27/2/4 | 0.0131 | 0.0122 | 4 | 145.48 |
| 264 | 4 | 7 | $17x+ 198x^2$ | 18 | 23/1/1 | 0.1865 | 0.2183 | $31x+ 66x^2$ | 18 | 24/2/4 | 0.1758 | 0.1813 | 4 | 1.20 |
| 272 | 4 | 7 | $33x+ 68x^2$ | 16 | 27/1/1 | 0.0367 | 0.0414 | $101x+204x^2$ | 16 | 27/2/2 | 0.0265 | 0.0310 | 2 | 1.34 |
| 280 | 4 | 7 | $103x+210x^2$ | 18 | 21/1/1 | 8.1004 | 11.396 | $17x+ 210x^2$ | 20 | 22/127 /254 | 8.2333 | 11.4211 | 4 | 1.00 |
| 288 | 4 | 7 | $19x+ 36x^2$ | 18 | 23/1/1 | 0.1190 | 0.1673 | $55x+ 72x^2$ | 18 | 25/1/3 | 0.0977 | 0.0895 | 4 | 1.87 |
| 296 | 4 | 5 | $19x+ 74x^2$ | 20 | 23/2/2 | 0.0749 | 0.1497 | $109x+ 222x^2$ | 20 | 24/2/4 | 0.0861 | 0.1186 | 4 | 1.26 |
| 304 | 4 | 5 | $37x+ 76x^2$ | 16 | 28/1/4 | 0.0163 | 0.0149 | $113x+ 76x^2$ | 16 | 28/1/4 | 0.0141 | 0.0122 | 4 | 1.22 |
| 312 | 4 | 5 | $19x+ 78x^2$ | 22 | 23/1/1 | 0.1109 | 0.1883 | $19x+ 78x^2$ | 22 | 23/1/1 | 0.1109 | 0.1883 | 4 | 1.00 |
| 320 | 4 | 5 | $21x+ 120x^2$ | 20 | 20/1/2 | 0.3842 | 0.6802 | $21x+ 80x^2$ | 20 | 25/1/3 | 0.0209 | 0.0283 | 4 | 24.04 |
| 328 | 4 | 5 | $21x+ 82x^2$ | 22 | 23/1/1 | 0.0252 | 0.0716 | $39x+ 246x^2$ | 22 | 27/4/8 | 0.0150 | 0.0236 | 4 | 3.03 |
| 336 | 3.5 | 5 | $115x+ 84x^2$ | 16 | 25/2/4 | 0.5643 | 0.7755 | $125x+252x^2$ | 16 | 28/1/4 | 0.3215 | 0.5000 | 2 | 1.55 |
| 344 | 3.5 | 5 | $193x+ 86x^2$ | 24 | 25/1/1 | 0.0605 | 0.1181 | $21x+ 258x^2$ | 24 | 25/1/1 | 0.0605 | 0.1181 | 4 | 1.00 |
| 352 | 3.5 | 5 | $21x+ 44x^2$ | 22 | 20/1/2 | 1.1520 | 2.0785 | $153x+264x^2$ | 22 | 27/1/1 | 0.0291 | 0.0381 | 2 | 54.55 |
| 360 | 3.5 | 5 | $133x+ 90x^2$ | 24 | 22/1/2 | 0.2943 | 0.4884 | $133x+ 90x^2$ | 24 | 22/1/2 | 0.2943 | 0.4884 | 4 | 1.00 |
| 368 | 3.5 | 5 | $81x+ 46x^2$ | 14 | 22/2/4 | 0.4270 | 0.7361 | $49x+ 276x^2$ | 20 | 20/1/2 | 1.0308 | 1.7973 | 4 | 0.41 |
| 376 | 3.5 | 5 | $45x+ 94x^2$ | 24 | 25/1/1 | 0.0437 | 0.0920 | $45x+ 94x^2$ | 24 | 25/1/1 | 0.0437 | 0.0920 | 4 | 1.00 |
| 384 | 3 | 5 | $23x+ 48x^2$ | 24 | 22/1/2 | 1.0699 | 2.4172 | $25x+ 336x^2$ | 24 | 25/1/3 | 0.6408 | 1.0269 | 4 | 2.35 |
| 392 | 3 | 5 | $243x+ 98x^2$ | 24 | 25/1/1 | 1.2396 | 2.4955 | $47x+ 98x^2$ | 24 | 27/1/3 | 1.2065 | 2.3457 | 4 | 1.06 |
| 400 | 3 | 5 | $151x+ 40x^2$ | 16 | 19/1/1 | 1.1190 | 3.7777 | $47+ 100x^2$ | 20 | 24/1/2 | 0.8329 | 1.2787 | 4 | 2.95 |
| 408 | 3 | 5 | $155x+102x^2$ | 24 | 23/1/1 | 0.1678 | 0.5815 | $25x+ 306x^2$ | 24 | 27/2/4 | 0.1063 | 0.2099 | 4 | 2.77 |
| 416 | 3 | 5 | $25x+ 52x^2$ | 26 | 23/1/1 | 0.9586 | 1.8504 | $129+ 104x^2$ | 26 | 25/1/1 | 0.1133 | 0.3088 | 4 | 5.99 |
| 424 | 3 | 5 | $51x+ 106x^2$ | 24 | 24/1/2 | 0.2928 | 0.5601 | $157x+106x^2$ | 24 | 27/1/3 | 0.1220 | 0.2075 | 8 | 2.70 |
| 432 | 3 | 5 | $47x+ 72x^2$ | 24 | 22/1/2 | 1.0130 | 1.8961 | $49x+ 72x^2$ | 24 | 24/1/2 | 0.7035 | 1.1129 | 4 | 1.70 |
| 440 | 3 | 5 | $91x+ 110x^2$ | 20 | 27/1/1 | 0.0523 | 0.1111 | $53x+ 330x^2$ | 24 | 27/2/4 | 0.0815 | 0.1627 | 4 | 0.68 |
| 448 | 3 | 3 | $29x+ 168x^2$ | 28 | 22/105/ 210 | 34.639 | 77.621 | $139x+112x^2$ | 28 | 25/1/1 | 1.1863 | 2.7474 | 8 | 28.25 |
| 456 | 3 | 3 | $29x+ 114x^2$ | 24 | 23/1/1 | 0.1113 | 0.4657 | $55x+ 342x^2$ | 24 | 27/1/3 | 0.0402 | 0.0680 | 4 | 6.85 |
| 464 | 3 | 3 | $247x+ 58x^2$ | 16 | 28/3/6 | 0.1053 | 0.1920 | $61x+ 348x^2$ | 24 | 22/1/2 | 0.6332 | 1.6873 | 4 | 0.11 |
| 472 | 3 | 3 | $29x+ 118x^2$ | 24 | 27/2/4 | 0.0523 | 0.1271 | $147x+354x^2$ | 24 | 27/1/3 | 0.0435 | 0.0777 | 4 | 1.64 |
| 480 | 3 | 3 | $89x+ 180x^2$ | 30 | 26/2/4 | 0.1321 | 0.3745 | $209x+120x^2$ | 30 | 27/1/1 | 0.0220 | 0.0919 | 4 | 4.08 |
| 488 | 3 | 3 | $91x+ 122x^2$ | 24 | 27/2/4 | 0.0714 | 0.1743 | $181x+122x^2$ | 24 | 27/1/3 | 0.0440 | 0.0747 | 4 | 2.33 |
| 496 | 3 | 3 | $157x+ 62x^2$ | 18 | 24/1/2 | 0.2355 | 0.4767 | $21x+ 124x^2$ | 24 | 26/1/2 | 0.1471 | 0.3231 | 4 | 1.48 |
| 504 | 3 | 3 | $55x+ 84x^2$ | 28 | 29/1/1 | 1.5408 | 3.8857 | $197x+168x^2$ | 28 | 25/1/3 | 0.1288 | 0.2926 | 2 | 13.28 |
| 512 | 3 | 3 | $31x+ 64x^2$ | 32 | 27/1/1 | 0.0906 | 0.2295 | $33x+ 448x^2$ | 32 | 27/1/1 | 0.0489 | 0.1329 | 4 | 1.73 |
| 528 | 3 | 3 | $17x+ 66x^2$ | 18 | 23/1/1 | 0.2698 | 0.8499 | $35x+ 396x^2$ | 32 | 29/1/1 | 0.0305 | 0.0492 | 4 | 17.27 |
| 544 | 3 | 3 | $35x+ 68x^2$ | 32 | 27/1/1 | 0.0367 | 0.1283 | $101x+ 204x^2$ | 32 | 27/1/1 | 0.0268 | 0.1116 | 8 | 1.15 |
| 560 | 3 | 3 | $227x+ 420x^2$ | 22 | 29/1/1 | 0.0082 | 0.0307 | $23x+ 140x^2$ | 28 | 22/266/ 532 | 68.04 | 190.7 | 4 | 0.00 |
| 576 | 3 | 3 | $65x+ 96x^2$ | 32 | 25/1/3 | 0.1678 | 0.3336 | $179x+ 432x^2$ | 32 | 28/12 | 0.0240 | 0.0793 | 4 | 4.21 |
| 592 | 3 | 3 | $19x+ 74x^2$ | 20 | 23/1/1 | 0.1272 | 0.5910 | $71x+ 148x^2$ | 28 | 28/1/2 | 0.0302 | 0.0791 | 4 | 7.47 |
| 608 | 2.75 | 3 | $37x+ 76x^2$ | 32 | 30/1/2 | 0.0390 | 0.1261 | $37x+ 456x^2$ | 32 | 30/1/2 | 0.0134 | 0.0482 | 4 | 2.62 |
| 624 | 2.75 | 3 | $41x+234x^2$ | 22 | 23/1/1 | 0.3643 | 1.4958 | $181x+156x^2$ | 26 | 27/2/2 | 0.0546 | 0.2513 | 2 | 5.95 |
| 640 | 2.75 | 3 | $39x+80x^2$ | 32 | 31/4/8 | 0.0328 | 0.0965 | $119x+480x^2$ | 32 | 32/2/4 | 0.0208 | 0.0554 | 4 | 1.74 |
| 656 | 2.75 | 3 | $185x+82x^2$ | 22 | 29/1/3 | 0.0381 | 0.1053 | $27x+492x^2$ | 32 | 28/1/2 | 0.1159 | 0.2826 | 8 | 0.37 |
| 672 | 2.75 | 3 | $43x+252x^2$ | 32 | 27/1/1 | 1.1711 | 3.9473 | $127x+168x^2$ | 32 | 25/1/3 | 0.1488 | 0.4019 | 4 | 9.82 |
| 688 | 2.75 | 3 | $21x+86x^2$ | 24 | 23/1/1 | 0.1745 | 0.9186 | $29x+516x^2$ | 32 | 29/1/1 | 0.0153 | 0.0719 | 4 | 12.78 |
| 704 | 2.75 | 3 | $155x+44x^2$ | 22 | 22/2/4 | 1.1057 | 3.8430 | $219x+528x^2$ | 32 | 33/1/1 | 0.0021 | 0.0111 | 4 | 346.22 |
| 720 | 2.75 | 3 | $79x+120x^2$ | 36 | 31/2/6 | 0.0235 | 0.0586 | $41x+480x^2$ | 36 | 32/2/4 | 0.0162 | 0.0472 | 4 | 1.24 |
| 736 | 2.75 | 3 | $139x+92x^2$ | 32 | 27/1/1 | 0.0477 | 0.1853 | $45x+92x^2$ | 32 | 31/2/6 | 0.0135 | 0.0349 | 12 | 5.31 |
| 752 | 2.75 | 3 | $23x+94x^2$ | 26 | 23/1/1 | 0.1795 | 0.9793 | $91x+188x^2$ | 32 | 29/1/3 | 0.0201 | 0.0568 | 4 | 17.24 |
| 768 | 2.75 | 3 | $217x+48x^2$ | 24 | 24/2/4 | 0.4168 | 1.4476 | $47x+672x^2$ | 32 | 31/2/6 | 0.0133 | 0.0533 | 4 | 27.16 |
| 784 | 2.75 | 3 | $25x+98x^2$ | 26 | 27/1/1 | 1.1551 | 4.5595 | $163x+196x^2$ | 32 | 29/1/3 | 0.0176 | 0.0441 | 4 | 103.39 |
| 800 | 2.75 | 3 | $17x+80x^2$ | 32 | 31/3/9 | 0.0334 | 0.0950 | $149x+300x^2$ | 32 | 31/2/6 | 0.0124 | 0.0339 | 4 | 2.80 |
| 816 | 2.75 | 3 | $127x+102x^2$ | 26 | 24/1/2 | 0.1721 | 0.7021 | $239x+612x^2$ | 34 | 28/2/4 | 0.0681 | 0.2585 | 2 | 2.72 |
| 832 | 2.75 | 1 | $25x+52x^2$ | 26 | 23/1/1 | 0.0851 | 0.7081 | $51x+104x^2$ | 32 | 31/2/6 | 0.0105 | 0.0291 | 32 | 24.33 |
| 848 | 2.75 | 1 | $239x+106x^2$ | 28 | 29/2/6 | 0.0272 | 0.0769 | $25x+52x^2$ | 32 | 29/1/3 | 0.0136 | 0.0384 | 16 | 2.00 |
| 864 | 2.75 | 1 | $17x+48x^2$ | 32 | 28/2/4 | 0.0289 | 0.1248 | $49x+288x^2$ | 36 | 33/3/9 | 0.0057 | 0.0165 | 16 | 21.89 |
| 880 | 2.75 | 1 | $137x+110x^2$ | 28 | 25/1/1 | 0.0304 | 0.2675 | $27x+660x^2$ | 32 | 32/3/6 | 0.0061 | 0.0268 | 4 | 43.85 |
| 896 | 2.75 | 1 | $215x+112x^2$ | 18 | 31/2/6 | 0.0097 | 0.0290 | $55x+112x^2$ | 32 | 31/2/6 | 0.0097 | 0.0290 | 32 | 1.00 |
| 912 | 2.75 | 1 | $29x+114x^2$ | 30 | 29/1/3 | 0.0126 | 0.0383 | $37x+228x^2$ | 38 | 31/1/1 | 0.0016 | 0.0145 | 6 | 2.64 |
| 928 | 2.75 | 1 | $15x+58x^2$ | 16 | 28/2/4 | 0.0268 | 0.1244 | $57x+116x^2$ | 32 | 31/2/6 | 0.0094 | 0.0290 | 32 | 4.29 |
| 944 | 2.75 | 1 | $147x+118x^2$ | 30 | 27/2/4 | 0.0428 | 0.2020 | $175x+708x^2$ | 34 | 35/4/10 | 0.0022 | 0.0083 | 4 | 24.34 |
| 960 | 2.5 | 1 | $29x+60x^2$ | 30 | 26/1/2 | 0.0602 | 0.2887 | $199x+240x^2$ | 40 | 31/1/1 | 0.0030 | 0.0284 | 4 | 10.17 |
| 976 | 2.5 | 1 | $59x+122x^2$ | 32 | 30/2/4 | 0.0185 | 0.0902 | $303x+244x^2$ | 36 | 35/2/4 | 0.0018 | 0.0089 | 8 | 10.13 |
| 992 | 2.25 | 1 | $65x+124x^2$ | 22 | 22/2/4 | 1.1803 | 5.8544 | $25x+744x^2$ | 32 | 35/1/1 | 0.0028 | 0.0092 | 4 | 636.35 |
| 1008 | 2 | 1 | $55x+84x^2$ | 28 | 30/322/644 | 9.9076 | 49.935 | $293x+252x^2$ | 42 | 33/1/1 | 0.0043 | 0.0436 | 4 | 1145.30 |
| 1024 | 2 | 1 | $31x+64x^2$ | 32 | 27/1/1 | 0.0538 | 0.5510 | $123x+256x^2$ | 34 | 28/1/2 | 0.0705 | 0.3610 | 4 | 1.53 |
| 1056 | 2 | 1 | $17x+66x^2$ | 18 | 24/1/2 | 0.3706 | 1.9570 | $43x+264x^2$ | 44 | 33/1/1 | 0.0041 | 0.0435 | 4 | 44.99 |
| 1088 | 2 | 1 | $171x+204x^2$ | 34 | 28/1/2 | 0.0662 | 0.3602 | $293x+252x^2$ | 42 | 33/1/1 | 0.0043 | 0.0436 | 4 | 8.26 |
| 1120 | 2 | 1 | $67x+140x^2$ | 24 | 27/1/1 | 0.0490 | 0.5493 | $73x+840x^2$ | 32 | 33/1/1 | 0.0039 | 0.0434 | 4 | 12.66 |
| 1152 | 2 | 1 | $35x+72x^2$ | 36 | 28/1/2 | 0.0624 | 0.3595 | $47x+288x^2$ | 48 | 35/1/1 | 0.0016 | 0.0186 | 8 | 19.33 |
| 1184 | 2 | 1 | $19x+74x^2$ | 20 | 26/1/2 | 0.1414 | 0.8370 | $385x+296x^2$ | 38 | 31/1/1 | 0.0085 | 0.1010 | 4 | 8.29 |
| 1216 | 2 | 1 | $39x+76x^2$ | 38 | 32/2/4 | 0.0217 | 0.1322 | $37x+912x^2$ | 38 | 33/1/1 | 0.0036 | 0.0433 | 8 | 3.05 |
| 1248 | 2 | 1 | $19x+78x^2$ | 22 | 24/1/2 | 0.3121 | 1.9476 | $53x+312x^2$ | 48 | 29/1/3 | 0.0565 | 0.2349 | 4 | 8.29 |
| 1280 | 2 | 1 | $199x+240x^2$ | 40 | 31/1/3 | 0.0236 | 0.1007 | $39x+960x^2$ | 40 | 35/1/1 | 0.0015 | 0.0185 | 8 | 5.44 |
| 1312 | 2 | 1 | $21x+82x^2$ | 22 | 28/1/2 | 0.0546 | 0.3580 | $33x+328x^2$ | 40 | 35/1/3 | 0.0042 | 0.0185 | 4 | 19.35 |
| 1344 | 2 | 1 | $211x+252x^2$ | 42 | 29/1/3 | 0.0523 | 0.2344 | $281x+336x^2$ | 48 | 29/1/3 | 0.0523 | 0.2344 | 4 | 1.00 |
| 1376 | 2 | 1 | $21x+86x^2$ | 24 | 28/1/2 | 0.0520 | 0.3576 | $177x+344x^2$ | 40 | 32/1/2 | 0.0096 | 0.0658 | 4 | 5.43 |
| 1408 | 2 | 1 | $43x+88x^2$ | 44 | 31/1/3 | 0.0214 | 0.1004 | $307x+352x^2$ | 44 | 35/1/1 | 0.0013 | 0.0185 | 16 | 5.43 |
| 1440 | 2 | 1 | $149x+60x^2$ | 30 | 28/2/4 | 0.0992 | 0.7142 | $61x+360x^2$ | 48 | 29/1/3 | 0.0487 | 0.2339 | 4 | 3.05 |
| 1472 | 2 | 1 | $45x+92x^2$ | 46 | 31/1/3 | 0.0204 | 0.1003 | $45x+1288x^2$ | 46 | 36/1/2 | 0.0016 | 0.0121 | 8 | 8.29 |
| 1504 | 2 | 1 | $49x+846x^2$ | 26 | 25/1/3 | 0.2533 | 1.2698 | $183x+376x^2$ | 46 | 33/1/3 | 0.0086 | 0.0430 | 4 | 29.53 |

Legend

| | |
|---|---|
| ▨ | LS-QPP-TUB(FER)min with TUB(FER) significantly smaller than that of LTE-QPP |
| 🟨 | LS-QPP-TUB(FER)min with TUB(FER) greater than that of LTE-QPP |

TABLE II

LTE-QPP AND LS-QPP WITH MINIMUM TUB(FER) INTERLEAVERS (MORE EXTENSIVE SEARCH)

| | | | LTE- QPP Interleavers | | | | | LS-QPP- TUB(FER) min Interleavers | | | | | | |
|---|---|---|---|---|---|---|---|---|---|---|---|---|---|---|
| L | SNR [dB] | num dist | $\pi(x)$ | D | $d_{min}$ /$N_1$ /$w_1$ | TUB (BER) $*10^7$ | TUB (FER) $*10^5$ | $\pi(x)$ | D | $d_{min}$ /$N_1$ /$w_1$ | TUB (BER) $*10^7$ | TUB (FER) $*10^5$ | No. Pol. | FER_LTE/ $FER_{min}$ |
| 72 | 7.5 | 9 | $7x+18x^2$ | 6 | 15/1/1 | 0.0521 | 0.0236 | $5x+54x^2$ | 6 | 16/2/4 | 0.0539 | 0.0166 | 4 | 1.42 |
| 168 | 5 | 7 | $101x+84x^2$ | 12 | 17/1/1 | 0.2745 | 0.3644 | $55x+84x^2$ | 12 | 25/2/2 | 0.9308 | 0.2035 | 4 | 1.79 |
| 368 | 3.5 | 5 | $81x+46x^2$ | 14 | 22/2/4 | 0.4270 | 0.7361 | $45x+92x^2$ | 16 | 28/1/4 | 0.0454 | 0.0337 | 8 | 21.84 |
| 440 | 3 | 5 | $91x+110x^2$ | 20 | 27/1/1 | 0.0523 | 0.1111 | $91x+110x^2$ | 20 | 27/1/1 | 0.0523 | 0.1111 | 4 | 1.00 |
| 464 | 3 | 3 | $247x+58x^2$ | 16 | 28/3/6 | 0.1053 | 0.1920 | $97x+116x^2$ | 20 | 29/1/1 | 0.0130 | 0.0407 | 4 | 4.72 |
| 560 | 3 | 3 | $227x+420x^2$ | 22 | 29/1/1 | 0.0082 | 0.0307 | $37x+420x^2$ | 22 | 29/1/3 | 0.0157 | 0.0278 | 4 | 1.07 |
| 656 | 2.75 | 3 | $185x+82x^2$ | 22 | 29/1/3 | 0.0381 | 0.1053 | $43x+164x^2$ | 22 | 29/1/3 | 0.0288 | 0.0720 | 4 | 1.46 |

TABEL III

SEARCH TIMES FOR SEVERAL LS-QPP WITH MINIMUM TUB(FER) INTERLEAVERS

| | | | | | | Previous method time | | | Improved method time | | |
|---|---|---|---|---|---|---|---|---|---|---|---|
| L | SNR [dB] | num_dist | $\pi(x)$ | D | TUB (FER) $*10^5$ | hours | min | sec | hours | min | sec |
| 528 | 3 | 3 | $35x+396x^2$ | 32 | 0.0492 | 2 | 8 | 50 | 0 | 17 | 20 |
| 752 | 2.75 | 3 | $91x+188x^2$ | 32 | 0.0568 | 7 | 25 | 52 | 0 | 57 | 41 |
| 944 | 2.75 | 1 | $175x+708x^2$ | 34 | 0.0083 | 4 | 15 | 28 | 2 | 43 | 45 |
| 1152 | 2 | 1 | $47x+288x^2$ | 48 | 0.0186 | 23 | 9 | 38 | 6 | 3 | 28 |
| 1472 | 2 | 1 | $45x+1288x^2$ | 46 | 0.0121 | too long | | | 78 | 51 | 15 |

TABLE IV

LTE-QPP AND $D_{LTE}$-QPP WITH MINIMUM TUB(FER) INTERLEAVERS

| | | | | | LTE- QPP Interleavers | | | | | $D_{LTE}$-QPP- TUB(FER)min Interleavers | | | | | | |
|---|---|---|---|---|---|---|---|---|---|---|---|---|---|---|---|---|
| L | SNR [dB] | num dist | $D_{LTE}$ | $D_{max}$ | $\pi(x)$ | ζ' | $d_{min}$ /$N_1$ /$w_1$ | TUB (BER) $*10^7$ | TUB (FER) $*10^5$ | $\pi(x)$ | ζ' | $d_{min}$ /$N_1$ /$w_1$ | TUB (BER) $*10^7$ | TUB (FER) $*10^5$ | No. pol. | FER_LTE/ $FER_{min}$ |
| 40 | 7.5 | 9 | 4 | 4 | $3x+10x^2$ | 2 | 11/1/3 | 10.559 | 1.6211 | $13x+30x^2$ | 2 | 12/1/2 | 4.0451 | 0.6539 | 4 | 2.48 |
| 48 | 7.5 | 9 | 6 | 6 | $7x+12x^2$ | 2 | 13/1/3 | 1.1890 | 0.1838 | $7x+36x^2$ | 2 | 15/2/6 | 0.7589 | 0.1150 | 2 | 1.60 |
| 56 | 7.5 | 9 | 6 | 6 | $19x+42x^2$ | 2 | 13/1/1 | 3.3169 | 0.9523 | $3x+42x^2$ | 2 | 13/1/1 | 3.3169 | 0.9523 | 4 | 1.00 |
| 64 | 7.5 | 9 | 8 | 8 | $7x+16x^2$ | 2 | 12/1/2 | 1.0947 | 0.3523 | $9x+48x^2$ | 2 | 12/1/2 | 1.1002 | 0.3456 | 4 | 1.02 |
| 72 | 7.5 | 9 | 6 | 8 | $7x+18x^2$ | 2 | 15/1/1 | 0.0521 | 0.0236 | $7x+60x^2$ | 3 | 15/1/3 | 0.4207 | 0.1007 | 4 | 0.23 |
| 80 | 6.5 | 9 | 10 | 10 | $11x+20x^2$ | 2 | 19/3/5 | 0.1369 | 0.0344 | $11x+20x^2$ | 2 | 19/3/5 | 0.1369 | 0.0344 | 4 | 1.00 |
| 88 | 6.5 | 9 | 8 | 8 | $5x+22x^2$ | 2 | 15/1/1 | 0.5231 | 0.2584 | $5x+22x^2$ | 2 | 15/1/1 | 0.5231 | 0.2584 | 4 | 1.00 |
| 96 | 6.5 | 9 | 12 | 12 | $11x+24x^2$ | 2 | 16/2/4 | 0.3015 | 0.1325 | $13x+72x^2$ | 2 | 16/1/2 | 0.2173 | 0.0942 | 4 | 1.41 |
| 104 | 6 | 9 | 8 | 8 | $7x+26x^2$ | 2 | 16/1/2 | 0.3928 | 0.2403 | $37x+26x^2$ | 2 | 16/1/2 | 0.4179 | 0.2028 | 4 | 1.18 |
| 112 | 6 | 9 | 14 | 14 | $41x+84x^2$ | 2 | 17/2/2 | 0.3613 | 0.2200 | $41x+28x^2$ | 2 | 17/2/2 | 0.3613 | 0.2200 | 4 | 1.00 |
| 120 | 6 | 7 | 10 | 10 | $103x+90x^2$ | 2 | 16/1/2 | 0.3376 | 0.2351 | $17x+90x^2$ | 2 | 16/1/2 | 0.3045 | 0.1609 | 4 | 1.46 |
| 128 | 5.5 | 7 | 16 | 16 | $15x+32x^2$ | 2 | 16/12 | 1.2349 | 0.6560 | $17x+32x^2$ | 2 | 18/1/2 | 0.2189 | 0.1446 | 4 | 4.54 |
| 136 | 5.5 | 7 | 10 | 10 | $9x+34x^2$ | 2 | 16/3/4 | 1.6234 | 1.0085 | $19x+102x^2$ | 2 | 16/1/2 | 0.9306 | 0.5515 | 4 | 1.83 |
| 144 | 5 | 7 | 16 | 16 | $17x+108x^2$ | 2 | 20/2/4 | 0.4829 | 0.2873 | $41x+24x^2$ | 3 | 18/2/4 | 0.7265 | 0.5425 | 4 | 0.53 |
| 152 | 5 | 7 | 12 | 12 | $9x+38x^2$ | 2 | 15/1/1 | 2.9059 | 2.6074 | $59x+38x^2$ | 2 | 16/1/2 | 2.2269 | 1.4680 | 4 | 1.78 |
| 160 | 5 | 7 | 16 | 16 | $21x+120x^2$ | 2 | 19/1/1 | 0.2742 | 0.2535 | $19x+120x^2$ | 2 | 19/1//1 | 0.2383 | 0.2274 | 4 | 1.11 |
| 168 | 5 | 7 | 12 | 12 | $101x+84x^2$ | 1 | 17/1/1 | 0.2745 | 0.3644 | $61x+126x^2$ | 2 | 21/1/1 | 1.4998 | 1.2596 | 4 | 0.29 |
| 176 | 5 | 7 | 16 | 16 | $21x+44x^2$ | 2 | 20/2/4 | 0.2018 | 0.1691 | $21x+44x^2$ | 2 | 20/2/4 | 0.2018 | 0.1691 | 2 | 1.00 |
| 184 | 5 | 7 | 12 | 14 | $57x+46x^2$ | 2 | 16/1/2 | 1.0900 | 0.9958 | $35x+138x^2$ | 2 | 18/1/2 | 0.4751 | 0.3942 | 4 | 2.53 |
| 192 | 4.5 | 7 | 16 | 16 | $23x+48x^2$ | 2 | 22/1/2 | 0.2096 | 0.1839 | $23x+144x^2$ | 2 | 22/1/2 | 0.1878 | 0.1735 | 4 | 1.06 |
| 200 | 4.5 | 7 | 14 | 14 | $13x+50x^2$ | 2 | 20/1/2 | 0.5008 | 0.4360 | $13x+150x^2$ | 2 | 20/1/2 | 0.4798 | 0.3839 | 4 | 1.14 |
| 208 | 4.5 | 7 | 16 | 16 | $27x+52x^2$ | 2 | 23/1/1 | 0.1889 | 0.1546 | $25x+52x^2$ | 2 | 23/1/1 | 0.1889 | 0.1546 | 4 | 1.00 |
| 216 | 4.5 | 7 | 18 | 18 | $11x+36x^2$ | 3 | 22/1/2 | 0.1029 | 0.1210 | $83x+36x^2$ | 3 | 23/1/1 | 0.1097 | 0.1061 | 4 | 1.14 |
| 224 | 4.5 | 7 | 16 | 16 | $27x+56x^2$ | 2 | 22/99 /198 | 2.5241 | 2.8293 | $27x+168x^2$ | 2 | 22/98 /196 | 2.5039 | 2.8129 | 4 | 1.01 |
| 232 | 4.5 | 7 | 16 | 16 | $85x+58x^2$ | 2 | 23/1/1 | 0.0226 | 0.0297 | $15x+174x^2$ | 2 | 23/1/1 | 0.0252 | 0.0295 | 4 | 1.01 |
| 240 | 4.5 | 7 | 16 | 16 | $29x+60x^2$ | 2 | 24/2/4 | 0.5208 | 0.4262 | $89x+60x^2$ | 2 | 24/1/2 | 0.0897 | 0.0807 | 4 | 5.28 |
| 248 | 4.5 | 7 | 18 | 18 | $33x+62x^2$ | 2 | 23/2/4 | 0.0451 | 0.0629 | $33x+186x^2$ | 2 | 23/1/1 | 0.0336 | 0.0406 | 4 | 1.55 |
| 256 | 4.5 | 7 | 16 | 16 | $15x+32x^2$ | 3 | 16/1/2 | 1.3491 | 1.7748 | $49x+160x^2$ | 3 | 20/2/4 | 0.3136 | 0.3524 | 4 | 5.04 |
| 264 | 4 | 7 | 18 | 18 | $17x+198x^2$ | 2 | 23/1/1 | 0.1865 | 0.2183 | $31x+66x^2$ | 2 | 24/2/4 | 0.1758 | 0.1813 | 4 | 1.20 |
| 272 | 4 | 7 | 16 | 16 | $33x+68x^2$ | 2 | 27/1/1 | 0.0367 | 0.0414 | $101x+204x^2$ | 2 | 27/2/2 | 0.0265 | 0.0310 | 2 | 1.34 |
| 280 | 4 | 7 | 18 | 20 | $103x+210x^2$ | 2 | 21/1/1 | 8.1004 | 11.396 | $87x+210x^2$ | 2 | 21/1/1 | 8.1004 | 11.396 | 4 | 1.00 |
| 288 | 4 | 7 | 18 | 18 | $19x+36x^2$ | 3 | 23/1/1 | 0.1190 | 0.1673 | $19x+108x^2$ | 3 | 23/1/1 | 0.1125 | 0.1637 | 4 | 1.02 |
| 296 | 4 | 5 | 20 | 20 | $19x+74x^2$ | 2 | 23/2/2 | 0.0749 | 0.1497 | $109x+222x^2$ | 2 | 24/2/4 | 0.0861 | 0.1186 | 4 | 1.26 |
| 304 | 4 | 5 | 16 | 16 | $37x+76x^2$ | 2 | 28/1/4 | 0.0163 | 0.0149 | $113x+76x^2$ | 2 | 28/1/4 | 0.0141 | 0.0122 | 4 | 1.22 |
| 312 | 4 | 5 | 22 | 22 | $19x+78x^2$ | 2 | 23/1/1 | 0.1109 | 0.1883 | $19x+78x^2$ | 2 | 23/1/1 | 0.1109 | 0.1883 | 4 | 1.00 |
| 320 | 4 | 5 | 20 | 20 | $21x+120x^2$ | 3 | 20/1/2 | 0.3842 | 0.6802 | $21x+40x^2$ | 3 | 20/1/2 | 0.4281 | 0.6706 | 8 | 1.01 |
| 328 | 4 | 5 | 22 | 22 | $21x+82x^2$ | 2 | 23/1/1 | 0.0252 | 0.0716 | $39x+246x^2$ | 2 | 27/4/8 | 0.0150 | 0.0236 | 4 | 3.03 |
| 336 | 3.5 | 5 | 16 | 16 | $115x+84x^2$ | 2 | 25/2/4 | 0.5643 | 0.7755 | $125x+252x^2$ | 2 | 28/1/4 | 0.3215 | 0.5000 | 2 | 1.55 |
| 344 | 3.5 | 5 | 24 | 24 | $193x+86x^2$ | 2 | 25/1/1 | 0.0605 | 0.1181 | $21x+258x^2$ | 2 | 25/1/1 | 0.0605 | 0.1181 | 4 | 1.00 |
| 352 | 3.5 | 5 | 22 | 22 | $21x+44x^2$ | 3 | 20/1/2 | 1.1520 | 2.0785 | $111+44x^2$ | 3 | 22/4/8 | 0.8100 | 1.4008 | 2 | 1.48 |
| 360 | 3.5 | 5 | 24 | 24 | $133x+90x^2$ | 2 | 22/1/2 | 0.2943 | 0.4884 | $133x+90x^2$ | 2 | 22/1/2 | 0.2943 | 0.4884 | 4 | 1.00 |
| 368 | 3.5 | 5 | 14 | 20 | $81x+46x^2$ | 3 | 22/2/4 | 0.4270 | 0.7361 | $25x+138x^2$ | 3 | 22/2/4 | 0.4270 | 0.7361 | 4 | 1.00 |
| 376 | 3.5 | 5 | 24 | 24 | $45x+94x^2$ | 2 | 25/1/1 | 0.0437 | 0.0920 | $45x+94x^2$ | 2 | 25/1/1 | 0.0437 | 0.0920 | 4 | 1.00 |
| 384 | 3 | 5 | 24 | 24 | $23x+48x^2$ | 3 | 22/1/2 | 1.0699 | 2.4172 | $25x+336x^2$ | 3 | 25/1/3 | 0.6408 | 1.0269 | 4 | 2.35 |
| 392 | 3 | 5 | 24 | 24 | $243x+98x^2$ | 2 | 25/1/1 | 1.2396 | 2.4955 | $47x+98x^2$ | 2 | 27/1/3 | 1.2065 | 2.3457 | 4 | 1.06 |
| 400 | 3 | 5 | 16 | 20 | $151x+40x^2$ | 5 | 19/1/1 | 1.1190 | 3.7777 | $151+120x^2$ | 5 | 26/1/1 | 0.1994 | 0.3832 | 4 | 9.86 |
| 408 | 3 | 5 | 24 | 24 | $155x+102x^2$ | 2 | 23/1/1 | 0.1678 | 0.5815 | $25x+306x^2$ | 2 | 27/2/4 | 0.1063 | 0.2099 | 4 | 2.77 |
| 416 | 3 | 5 | 26 | 26 | $25x+52x^2$ | 3 | 23/1/1 | 0.9586 | 1.8504 | $27+156x^2$ | 3 | 24/1/2 | 0.3624 | 0.6737 | 8 | 2.75 |
| 424 | 3 | 5 | 24 | 24 | $51x+106x^2$ | 2 | 24/1/2 | 0.2928 | 0.5601 | $157x+106x^2$ | 2 | 27/1/3 | 0.1220 | 0.2075 | 8 | 2.70 |
| 432 | 3 | 5 | 24 | 24 | $47x+72x^2$ | 3 | 22/1/2 | 1.0130 | 1.8961 | $49x+72x^2$ | 3 | 24/1/2 | 0.7035 | 1.1129 | 4 | 1.70 |
| 440 | 3 | 5 | 20 | 24 | $91x+110x^2$ | 2 | 27/1/1 | 0.0523 | 0.1111 | $91x+110x^2$ | 2 | 27/1/1 | 0.0523 | 0.1111 | 4 | 1.00 |
| 448 | 3 | 3 | 28 | 28 | $29x+168x^2$ | 3 | 22/105/ 210 | 34.639 | 77.621 | $29x+56x^2$ | 3 | 22/105/ 210 | 34.092 | 76.374 | 12 | 1.02 |
| 456 | 3 | 3 | 24 | 24 | $29x+114x^2$ | 2 | 23/1/1 | 0.1113 | 0.4657 | $55x+342x^2$ | 2 | 27/1/3 | 0.0402 | 0.0680 | 4 | 6.85 |
| 464 | 3 | 3 | 16 | 24 | $247x+58x^2$ | 3 | 28/3/6 | 0.1053 | 0.1920 | $15x+406x^2$ | 3 | 28/2/4 | 0.0652 | 0.1407 | 4 | 1.36 |
| 472 | 3 | 3 | 24 | 24 | $29x+118x^2$ | 2 | 27/2/4 | 0.0523 | 0.1271 | $147x+354x^2$ | 2 | 27/1/3 | 0.0435 | 0.0777 | 4 | 1.64 |
| 480 | 3 | 3 | 30 | 30 | $89x+180x^2$ | 3 | 26/2/4 | 0.1321 | 0.3745 | $151x+180x^2$ | 3 | 26/1/2 | 0.0949 | 0.2564 | 4 | 1.46 |
| 488 | 3 | 3 | 24 | 24 | $91x+122x^2$ | 2 | 27/2/4 | 0.0714 | 0.1743 | $181x+122x^2$ | 2 | 27/1/3 | 0.0440 | 0.0747 | 4 | 2.33 |
| 496 | 3 | 3 | 18 | 24 | $157x+62x^2$ | 3 | 24/1/2 | 0.2355 | 0.4767 | $109x+186x^2$ | 3 | 26/1/2 | 0.1147 | 0.2212 | 4 | 2.16 |
| 504 | 3 | 3 | 28 | 28 | $55x+84x^2$ | 3 | 29/1/1 | 1.5408 | 3.8857 | $55x+84x^2$ | 3 | 29/1/1 | 1.5408 | 3.8857 | 4 | 1.00 |
| 512 | 3 | 3 | 32 | 32 | $31x+64x^2$ | 3 | 27/1/1 | 0.0906 | 0.2295 | $33x+448x^2$ | 3 | 27/1/1 | 0.0489 | 0.1329 | 4 | 1.73 |
| 528 | 3 | 3 | 18 | 32 | $17x+66x^2$ | 3 | 23/1/1 | 0.2698 | 0.8499 | $17x+198x^2$ | 3 | 24/1/2 | 0.2237 | 0.5120 | 16 | 1.66 |
| 544 | 3 | 3 | 32 | 32 | $35x+68x^2$ | 3 | 27/1/1 | 0.0367 | 0.1283 | $101x+204x^2$ | 3 | 27/1/1 | 0.0268 | 0.1116 | 8 | 1.15 |
| 560 | 3 | 3 | 22 | 28 | $227x+420x^2$ | 2 | 29/1/1 | 0.0082 | 0.0307 | $37x+420x^2$ | 2 | 29/1/3 | 0.0157 | 0.0278 | 4 | 1.10 |
| 576 | 3 | 3 | 32 | 32 | $65x+96x^2$ | 3 | 25/1/3 | 0.1678 | 0.3336 | $35x+72x^2$ | 3 | 28/12 | 0.0326 | 0.1040 | 4 | 3.21 |
| 592 | 3 | 3 | 20 | 28 | $19x+74x^2$ | 3 | 23/1/1 | 0.1272 | 0.5910 | $129x+74x^2$ | 3 | 25/1/1 | 0.0866 | 0.3059 | 4 | 1.93 |
| 608 | 2.75 | 3 | 32 | 32 | $37x+76x^2$ | 3 | 30/1/2 | 0.0390 | 0.1261 | $265x+380x^2$ | 3 | 31/3/7 | 0.0324 | 0.0856 | 4 | 1.47 |
| 624 | 2.75 | 3 | 22 | 26 | $41x+234x^2$ | 3 | 23/1/1 | 0.3643 | 1.4958 | $197x+546x^2$ | 3 | 26/2/4 | 0.2816 | 0.7612 | 4 | 1.97 |
| 640 | 2.75 | 3 | 32 | 32 | $39x+80x^2$ | 3 | 31/4/8 | 0.0328 | 0.0965 | $39x+560x^2$ | 3 | 31/2/6 | 0.0365 | 0.0908 | 8 | 1.06 |
| 656 | 2.75 | 3 | 22 | 32 | $185x+82x^2$ | 3 | 29/1/3 | 0.0381 | 0.1053 | $21x+246x^2$ | 3 | 29/1/3 | 0.0348 | 0.1019 | 4 | 1.03 |
| 672 | 2.75 | 3 | 32 | 32 | $43x+252x^2$ | 3 | 27/1/1 | 1.1711 | 3.9473 | $125x+588x^2$ | 3 | 30/157/ 314 | 1.1381 | 3.8221 | 4 | 1.03 |
| 688 | 2.75 | 3 | 24 | 32 | $21x+86x^2$ | 3 | 23/1/1 | 0.1745 | 0.9186 | $21x+430x^2$ | 3 | 29/1/1 | 0.0394 | 0.1403 | 8 | 6.55 |
| 704 | 2.75 | 3 | 22 | 32 | $155x+44x^2$ | 4 | 22/2/4 | 1.1057 | 3.8430 | $111x+132x^2$ | 4 | 25/2/6 | 0.3609 | 1.0001 | 2 | 3.84 |
| 720 | 2.75 | 3 | 36 | 36 | $79x+120x^2$ | 3 | 31/2/6 | 0.0235 | 0.0586 | $79x+120x^2$ | 3 | 31/2/6 | 0.0235 | 0.0586 | 4 | 1.00 |
| 736 | 2.75 | 3 | 32 | 32 | $139x+92x^2$ | 3 | 27/1/1 | 0.0477 | 0.1853 | $45x+92x^2$ | 3 | 31/2/6 | 0.0135 | 0.0349 | 8 | 5.31 |
| 752 | 2.75 | 3 | 26 | 32 | $23x+94x^2$ | 3 | 23/1/1 | 0.1795 | 0.9793 | $165x+94x^2$ | 3 | 26/1/2 | 0.1560 | 0.5092 | 4 | 1.92 |
| 768 | 2.75 | 3 | 24 | 32 | $217x+48x^2$ | 4 | 24/2/4 | 0.4168 | 1.4476 | $217x+144x^2$ | 4 | 26/2/6 | 0.1824 | 0.5982 | 4 | 2.42 |
| 784 | 2.75 | 3 | 26 | 32 | $25x+98x^2$ | 3 | 27/1/1 | 1.1551 | 4.5595 | $25x+294x^2$ | 3 | 29/1/3 | 1.1381 | 4.4394 | 4 | 1.03 |
| 800 | 2.75 | 3 | 32 | 32 | $17x+80x^2$ | 5 | 31/3/9 | 0.0334 | 0.0950 | $17x+80x^2$ | 5 | 31/3/9 | 0.0334 | 0.0950 | 4 | 1.00 |
| 816 | 2.75 | 3 | 26 | 34 | $127x+102x^2$ | 3 | 24/1/2 | 0.1721 | 0.7021 | $53x+510x^2$ | 3 | 28/2/4 | 0.0964 | 0.3118 | 8 | 2.25 |
| 832 | 2.75 | 1 | 26 | 32 | $25x+52x^2$ | 4 | 23/1/1 | 0.0851 | 0.7081 | $27x+572x^2$ | 4 | 27/1/3 | 0.0366 | 0.1016 | 16 | 6.97 |
| 848 | 2.75 | 1 | 28 | 32 | $239x+106x^2$ | 3 | 29/2/6 | 0.0272 | 0.0769 | $185x+318x^2$ | 3 | 29/1/1 | 0.0045 | 0.0384 | 4 | 2.00 |
| 864 | 2.75 | 1 | 32 | 36 | $17x+48x^2$ | 8 | 28/2/4 | 0.0289 | 0.1248 | $113x+48x^2$ | 8 | 28/1/2 | 0.0144 | 0.0624 | 16 | 2.00 |
| 880 | 2.75 | 1 | 28 | 32 | $137x+110x^2$ | 3 | 25/1/1 | 0.0304 | 0.2675 | $53x+770x^2$ | 3 | 32/1/2 | 0.0020 | 0.0089 | 4 | 30.06 |
| 896 | 2.75 | 1 | 18 | 32 | $215x+112x^2$ | 3 | 31/2/6 | 0.0097 | 0.0290 | $215x+112x^2$ | 3 | 31/2/6 | 0.0097 | 0.0290 | 4 | 1.00 |
| 912 | 2.75 | 1 | 30 | 38 | $29x+114x^2$ | 3 | 29/1/3 | 0.0126 | 0.0383 | $283x+570x^2$ | 3 | 33/1/1 | 0.0006 | 0.0055 | 4 | 6.96 |
| 928 | 2.75 | 1 | 16 | 32 | $15x+58x^2$ | 4 | 28/2/4 | 0.0268 | 0.1244 | $85x+522x^2$ | 4 | 30/2/4 | 0.0102 | 0.0471 | 4 | 2.64 |
| 944 | 2.75 | 1 | 30 | 34 | $147x+118x^2$ | 3 | 27/2/4 | 0.0428 | 0.2020 | $147x+354x^2$ | 3 | 31/1/1 | 0.0015 | 0.0145 | 8 | 13.93 |
| 960 | 2.5 | 1 | 30 | 40 | $29x+60x^2$ | 4 | 26/1/2 | 0.0602 | 0.2887 | $89x+300x^2$ | 4 | 28/1/2 | 0.0238 | 0.1142 | 16 | 2.53 |
| 976 | 2.5 | 1 | 32 | 36 | $59x+122x^2$ | 3 | 30/2/4 | 0.0185 | 0.0902 | $181x+122x^2$ | 3 | 32/1/2 | 0.0037 | 0.0178 | 8 | 5.07 |
| 992 | 2.25 | 1 | 22 | 32 | $65x+124x^2$ | 3 | 22/2/4 | 1.1803 | 5.8544 | $65x+124x^2$ | 3 | 22/2/4 | 1.1803 | 5.8544 | 16 | 1.00 |
| 1008 | 2 | 1 | 28 | 42 | $55x+84x^2$ | 4 | 30/322/ 644 | 9.9076 | 49.935 | $55x+588x^2$ | 4 | 29/1/1 | 0.0235 | 0.2367 | 4 | 210.96 |
| 1024 | 2 | 1 | 32 | 34 | $31x+64x^2$ | 4 | 27/1/1 | 0.0538 | 0.5510 | $31x+192x^2$ | 4 | 29/1/1 | 0.0231 | 0.2365 | 20 | 2.33 |
| 1056 | 2 | 1 | 18 | 44 | $17x+66x^2$ | 4 | 24/1/2 | 0.3706 | 1.9570 | $17x+330x^2$ | 4 | 26/1/2 | 0.1591 | 0.8400 | 20 | 2.33 |
| 1088 | 2 | 1 | 34 | 42 | $171x+204x^2$ | 4 | 28/1/2 | 0.0662 | 0.3602 | $35x+748x^2$ | 4 | 29/1/1 | 0.0651 | 0.2360 | 8 | 1.53 |
| 1120 | 2 | 1 | 24 | 32 | $67x+140x^2$ | 3 | 27/1/1 | 0.0490 | 0.5493 | $117x+980x^2$ | 3 | 35/1/3 | 0.0050 | 0.0186 | 4 | 29.53 |
| 1152 | 2 | 1 | 36 | 48 | $35x+72x^2$ | 4 | 28/1/2 | 0.0624 | 0.3595 | $181x+216x^2$ | 4 | 30/1/2 | 0.0268 | 0.1543 | 4 | 2.33 |
| 1184 | 2 | 1 | 20 | 38 | $19x+74x^2$ | 4 | 26/1/2 | 0.1414 | 0.8370 | $129x+74x^2$ | 4 | 28/1/2 | 0.0607 | 0.3592 | 6 | 2.33 |
| 1216 | 2 | 1 | 38 | 38 | $39x+76x^2$ | 4 | 32/2/4 | 0.0217 | 0.1322 | $39x+684x^2$ | 4 | 32/1/2 | 0.0109 | 0.0661 | 16 | 2..00 |
| 1248 | 2 | 1 | 22 | 48 | $19x+78x^2$ | 4 | 24/1/2 | 0.3121 | 1.9476 | $19x+234x^2$ | 4 | 26/1/2 | 0.1339 | 0.8357 | 38 | 2.33 |
| 1280 | 2 | 1 | 40 | 40 | $199x+240x^2$ | 4 | 31/1/3 | 0.0236 | 0.1007 | $41x+560x^2$ | 4 | 32/1/2 | 0.0103 | 0.0660 | 24 | 1.53 |
| 1312 | 2 | 1 | 22 | 40 | $21x+82x^2$ | 4 | 28/1/2 | 0.0546 | 0.3580 | $21x+1230x^2$ | 4 | 32/1/2 | 0.0101 | 0.0659 | 2 | 5.43 |
| 1344 | 2 | 1 | 42 | 48 | $211x+252x^2$ | 4 | 29/1/3 | 0.0523 | 0.2344 | $43x+252x^2$ | 4 | 29/1/3 | 0.0523 | 0.2344 | 64 | 1.00 |
| 1376 | 2 | 1 | 24 | 40 | $21x+86x^2$ | 4 | 28/1/2 | 0.0520 | 0.3576 | $151x+946x^2$ | 4 | 30/1/2 | 0.0223 | 0.1534 | 10 | 2.33 |
| 1408 | 2 | 1 | 44 | 44 | $43x+88x^2$ | 4 | 31/1/3 | 0.0214 | 0.1004 | $45x+968x^2$ | 4 | 33/1/3 | 0.0092 | 0.0431 | 40 | 2.33 |
| 1440 | 2 | 1 | 30 | 48 | $149x+60x^2$ | 6 | 28/2/4 | 0.0992 | 0.7142 | $151x+300x^2$ | 6 | 32/1/4 | 0.0183 | 0.0657 | 12 | 10.87 |
| 1472 | 2 | 1 | 46 | 46 | $45x+92x^2$ | 4 | 31/1/3 | 0.0204 | 0.1003 | $47x+92x^2$ | 4 | 33/1/3 | 0.0088 | 0.0430 | 60 | 2.33 |
| 1504 | 2 | 1 | 26 | 46 | $49x+846x^2$ | 4 | 25/1/3 | 0.2533 | 1.2698 | $23x+1410x^2$ | 4 | 27/1/3 | 0.1086 | 0.5446 | 32 | 2.33 |

Legend (for lengths from 40 up to 512):

| | |
|---|---|
| ▨ | Better than LTE-QPP and as good as LS-QPP-TUB(FER)min |
| 🟥 | Better than LTE-QPP and better than LS-QPP-TUB(FER)min |
| 🟩 | As good as LTE-QPP and better than LS-QPP-TUB(FER)min |
| 🟢 | As good as LTE-QPP, but weaker than LS-QPP-TUB(FER)min |
| 🟦 | Better than LTE-QPP, but weaker than LS-QPP-TUB(FER)min |
| 🟪 | Weaker than LTE-QPP, but better than LS-QPP-TUB(FER)min |
| 🟫 | Weaker than LTE-QPPand weaker than LS-QPP-TUB(FER)min |
| 🟨 | Weaker than LTE-QPP but as weak as LS-QPP-TUB(FER)min |

TABLE V

LTE-QPP AND QPP WITH MINIMUM TUB(FER) INTERLEAVERS

| L | SNR [dB] | num dist | $D_{LTE}$ | $D_{opt}$ | LTE- QPP Interleavers $\pi(x)$ | $\zeta'$ | $d_{min}$ /$N_1$ /$w_1$ | TUB (BER) $*10^7$ | TUB (FER) $*10^5$ | QPP- TUB(FER)min Interleavers $\pi(x)$ | $\zeta'$ | $d_{min}$ /$N_1$ /$w_1$ | TUB (BER) $*10^7$ | TUB (FER) $*10^5$ | No. pol. | FER_LTE/ $FER_{min}$ |
|---|---|---|---|---|---|---|---|---|---|---|---|---|---|---|---|---|
| 40 | 7.5 | 9 | 4 | 4 | $3x+10x^2$ | 2 | 11/1/3 | 10.559 | 1.6211 | $13x+30x^2$ | 2 | 12/1/2 | 4.0451 | 0.6539 | 4 | 2.48 |
| 48 | 7.5 | 9 | 6 | 6 | $7x+12x^2$ | 2 | 13/1/3 | 1.1890 | 0.1838 | $7x+36x^2$ | 2 | 15/2/6 | 0.7589 | 0.1150 | 2 | 1.60 |
| 56 | 7.5 | 9 | 6 | 4 | $19x+42x^2$ | 2 | 13/1/1 | 3.3169 | 0.9523 | $5x+42x^2$ | 2 | 14/21/ 42 | 2.9899 | 0.8041 | 4 | 1.18 |
| 64 | 7.5 | 9 | 8 | 4 | $7x+16x^2$ | 2 | 12/1/2 | 1.0947 | 0.3523 | $19x+48x^2$ | 2 | 15/1/3 | 0.1317 | 0.0255 | 4 | 13.82 |
| 72 | 7.5 | 9 | 6 | 6 | $7x+18x^2$ | 2 | 15/1/1 | 0.0521 | 0.0236 | $5x+54x^2$ | 2 | 15/1/3 | 0.4207 | 0.0166 | 4 | 1.42 |
| 80 | 6.5 | 9 | 10 | 10 | $11x+20x^2$ | 2 | 19/3/5 | 0.1369 | 0.0344 | $11x+20x^2$ | 2 | 19/3/5 | 0.1369 | 0.0344 | 4 | 1.00 |
| 88 | 6.5 | 9 | 8 | 6 | $5x+22x^2$ | 2 | 15/1/1 | 0.5231 | 0.2584 | $13x+22x^2$ | 2 | 17/1/1 | 0.1754 | 0.0617 | 4 | 4.19 |
| 96 | 6.5 | 9 | 12 | 6 | $11x+24x^2$ | 2 | 16/2/4 | 0.3015 | 0.1325 | $5x+72x^2$ | 2 | 17/2/4 | 0.1263 | 0.0451 | 4 | 2.94 |
| 104 | 6 | 9 | 8 | 6 | $7x+26x^2$ | 2 | 16/1/2 | 0.3928 | 0.2403 | $31x+78x^2$ | 2 | 18/1/2 | 0.2909 | 0.1000 | 4 | 2.40 |
| 112 | 6 | 9 | 14 | 6 | $41x+84x^2$ | 2 | 17/2/2 | 0.3613 | 0.2200 | $17x+28x^2$ | 2 | 19/1/3 | 0.4649 | 0.1779 | 4 | 1.24 |
| 120 | 6 | 7 | 10 | 6 | $103x+90x^2$ | 2 | 16/1/2 | 0.3376 | 0.2351 | $11x+90x^2$ | 2 | 19/1/1 | 0.1309 | 0.0540 | 2 | 4.35 |
| 128 | 5.5 | 7 | 16 | 16 | $15x+32x^2$ | 2 | 16/12 | 1.2349 | 0.6560 | $17x+32x^2$ | 2 | 18/1/2 | 0.2189 | 0.1446 | 4 | 4.54 |
| 136 | 5.5 | 7 | 10 | 8 | $9x+34x^2$ | 2 | 16/3/4 | 1.6234 | 1.0085 | $47x+102x^2$ | 2 | 19/1/1 | 0.1913 | 0.0981 | 4 | 10.28 |
| 144 | 5 | 7 | 16 | 6 | $17x+108x^2$ | 2 | 20/2/4 | 0.4829 | 0.2873 | $11x+36x^2$ | 2 | 22/1/2 | 0.2675 | 0.1416 | 4 | 2.03 |
| 152 | 5 | 7 | 12 | 6 | $9x+38x^2$ | 2 | 15/1/1 | 2.9059 | 2.6074 | $23x+38x^2$ | 2 | 20/4/8 | 0.3913 | 0.2691 | 4 | 9.69 |
| 160 | 5 | 7 | 16 | 10 | $21x+120x^2$ | 2 | 19/1/1 | 0.2742 | 0.2535 | $9x+120x^2$ | 2 | 21/2/6 | 0.2894 | 0.1527 | 4 | 1.66 |
| 168 | 5 | 7 | 12 | 6 | $101x+84x^2$ | 1 | 17/1/1 | 0.2745 | 0.3644 | $13x+42x^2$ | 2 | 22/141/ 282 | 1.5090 | 1.2461 | 4 | 0.29 |
| 176 | 5 | 7 | 16 | 6 | $21x+44x^2$ | 2 | 20/2/4 | 0.2018 | 0.1691 | $9x+132x^2$ | 2 | 21/1/1 | 0.3467 | 0.1338 | 4 | 1.26 |
| 184 | 5 | 7 | 12 | 8 | $57x+46x^2$ | 2 | 16/1/2 | 1.0900 | 0.9958 | $27x+138x^2$ | 2 | 22/1/2 | 0.0532 | 0.0442 | 4 | 22.53 |
| 192 | 4.5 | 7 | 16 | 6 | $23x+48x^2$ | 2 | 22/1/2 | 0.2096 | 0.1839 | $79x+48x^2$ | 2 | 25/6/30 | 0.2943 | 0.1327 | 4 | 1.39 |
| 200 | 4.5 | 7 | 14 | 10 | $13x+50x^2$ | 2 | 20/1/2 | 0.5008 | 0.4360 | $11x+150x^2$ | 2 | 20/1/2 | 0.3513 | 0.3055 | 4 | 1.43 |
| 208 | 4.5 | 7 | 16 | 8 | $27x+52x^2$ | 2 | 23/1/1 | 0.1889 | 0.1546 | $85x+156x^2$ | 2 | 23/1/1 | 0.0887 | 0.0720 | 4 | 2.15 |
| 216 | 4.5 | 7 | 18 | 18 | $11x+36x^2$ | 3 | 22/1/2 | 0.1029 | 0.1210 | $23x+144x^2$ | 2 | 22/1/2 | 0.0925 | 0.1038 | 4 | 1.17 |
| 224 | 4.5 | 7 | 16 | 16 | $27x+56x^2$ | 2 | 22/99 /198 | 2.5241 | 2.8293 | $27x+168x^2$ | 2 | 22/98 /196 | 2.5039 | 2.8129 | 4 | 1.01 |
| 232 | 4.5 | 7 | 16 | 16 | $85x+58x^2$ | 2 | 23/1/1 | 0.0226 | 0.0297 | $15x+174x^2$ | 2 | 23/1/1 | 0.0252 | 0.0295 | 4 | 1.01 |
| 240 | 4.5 | 7 | 16 | 12 | $29x+60x^2$ | 2 | 24/2/4 | 0.5208 | 0.4262 | $71x+180x^2$ | 2 | 26/4/12 | 0.0502 | 0.0318 | 2 | 13.40 |
| 248 | 4.5 | 7 | 18 | 18 | $33x+62x^2$ | 2 | 23/2/4 | 0.0451 | 0.0629 | $33x+186x^2$ | 2 | 23/1/1 | 0.0336 | 0.0406 | 4 | 1.55 |
| 256 | 4.5 | 7 | 16 | 16 | $15x+32x^2$ | 3 | 16/1/2 | 1.3491 | 1.7748 | $31x+192x^2$ | 2 | 27/2/4 | 0.0131 | 0.0122 | 4 | 145.48 |
| 264 | 4 | 7 | 18 | 18 | $17x+198x^2$ | 2 | 23/1/1 | 0.1865 | 0.2183 | $31x+66x^2$ | 2 | 24/2/4 | 0.1758 | 0.1813 | 4 | 1.20 |
| 272 | 4 | 7 | 16 | 16 | $33x+68x^2$ | 2 | 27/1/1 | 0.0367 | 0.0414 | $101x+204x^2$ | 2 | 27/2/2 | 0.0265 | 0.0310 | 2 | 1.34 |
| 280 | 4 | 7 | 18 | 10 | $103x+210x^2$ | 2 | 21/1/1 | 8.1004 | 11.396 | $13x+70x^2$ | 2 | 22/126/ 252 | 8.0556 | 11.307 | 4 | 1.01 |
| 288 | 4 | 7 | 18 | 16 | $19x+36x^2$ | 3 | 23/1/1 | 0.1190 | 0.1673 | $35x+72x^2$ | 2 | 28/5/12 | 0.0285 | 0.0303 | 4 | 5.52 |
| 296 | 4 | 5 | 20 | 10 | $19x+74x^2$ | 2 | 23/2/2 | 0.0749 | 0.1497 | $41x+74x^2$ | 2 | 28/6/14 | 0.0366 | 0.0402 | 4 | 3.72 |
| 304 | 4 | 5 | 16 | 16 | $37x+76x^2$ | 2 | 28/1/4 | 0.0163 | 0.0149 | $113x+76x^2$ | 2 | 28/1/4 | 0.0141 | 0.0122 | 4 | 1.22 |
| 312 | 4 | 5 | 22 | 12 | $19x+78x^2$ | 2 | 23/1/1 | 0.1109 | 0.1883 | $43x+78x^2$ | 2 | 28/5/10 | 0.0162 | 0.0229 | 4 | 8.22 |
| 320 | 4 | 5 | 20 | 14 | $21x+120x^2$ | 3 | 20/1/2 | 0.3842 | 0.6802 | $43x+240x^2$ | 2 | 28/3/6 | 0.0163 | 0.0187 | 4 | 36.37 |
| 328 | 4 | 5 | 22 | 22 | $21x+82x^2$ | 2 | 23/1/1 | 0.0252 | 0.0716 | $39x+246x^2$ | 2 | 27/4/8 | 0.0150 | 0.0236 | 4 | 3.03 |
| 336 | 3.5 | 5 | 16 | 12 | $115x+84x^2$ | 2 | 25/2/4 | 0.5643 | 0.7755 | $55x+252x^2$ | 2 | 25/2/2 | 0.3587 | 0.4505 | 2 | 1.72 |
| 344 | 3.5 | 5 | 24 | 8 | $193x+86x^2$ | 2 | 25/1/1 | 0.0605 | 0.1181 | $27x+258x^2$ | 2 | 26/1/4 | 0.1328 | 0.1133 | 4 | 1.04 |
| 352 | 3.5 | 5 | 22 | 22 | $21x+44x^2$ | 3 | 20/1/2 | 1.1520 | 2.0785 | $153+264x^2$ | 2 | 27/1/1 | 0.0291 | 0.0381 | 2 | 54.55 |
| 360 | 3.5 | 5 | 24 | 18 | $133x+90x^2$ | 2 | 22/1/2 | 0.2943 | 0.4884 | $79x+120x^2$ | 2 | 30/2/8 | 0.0251 | 0.0277 | 4 | 17.63 |
| 368 | 3.5 | 5 | 14 | 16 | $81x+46x^2$ | 3 | 22/2/4 | 0.4270 | 0.7361 | $45x+92x^2$ | 2 | 28/1/4 | 0.0454 | 0.0337 | 8 | 21.84 |
| 376 | 3.5 | 5 | 24 | 12 | $45x+94x^2$ | 2 | 25/1/1 | 0.0437 | 0.0920 | $77x+282x^2$ | 2 | 29/3/11 | 0.0416 | 0.0407 | 4 | 2.26 |
| 384 | 3 | 5 | 24 | 16 | $23x+48x^2$ | 3 | 22/1/2 | 1.0699 | 2.4172 | $47x+96x^2$ | 2 | 28/1/4 | 0.1916 | 0.1466 | 8 | 16.49 |
| 392 | 3 | 5 | 24 | 6 | $243x+98x^2$ | 2 | 25/1/1 | 1.2396 | 2.4955 | $75x+210x^2$ | 8 | 25/2/4 | 0.6770 | 0.9476 | 4 | 2.63 |
| 400 | 3 | 5 | 16 | 16 | $151x+40x^2$ | 5 | 19/1/1 | 1.1190 | 3.7777 | $49+100x^2$ | 2 | 28/1/4 | 0.1830 | 0.1459 | 8 | 25.89 |
| 408 | 3 | 5 | 24 | 12 | $155x+102x^2$ | 2 | 23/1/1 | 0.1678 | 0.5815 | $31x+306x^2$ | 2 | 27/1/3 | 0.1424 | 0.1873 | 4 | 3.10 |
| 416 | 3 | 5 | 26 | 16 | $25x+52x^2$ | 3 | 23/1/1 | 0.9586 | 1.8504 | $51+104x^2$ | 3 | 28/1/4 | 0.1751 | 0.1452 | 8 | 12.74 |
| 424 | 3 | 5 | 24 | 14 | $51x+106x^2$ | 2 | 24/1/2 | 0.2928 | 0.5601 | $57x+318x^2$ | 2 | 31/2/6 | 0.0425 | 0.0555 | 4 | 10.09 |
| 432 | 3 | 5 | 24 | 18 | $47x+72x^2$ | 3 | 22/1/2 | 1.0130 | 1.8961 | $23x+360x^2$ | 3 | 30/2/8 | 0.0870 | 0.0826 | 16 | 22.96 |
| 440 | 3 | 5 | 20 | 16 | $91x+110x^2$ | 2 | 27/1/1 | 0.0523 | 0.1111 | $59x+110x^2$ | 2 | 27/1/1 | 0.0416 | 0.0985 | 4 | 1.13 |
| 448 | 3 | 3 | 28 | 16 | $29x+168x^2$ | 3 | 22/105/ 210 | 34.639 | 77.621 | $55x+112x^2$ | 3 | 28/1/4 | 0.1020 | 0.0929 | 8 | 835.53 |
| 456 | 3 | 3 | 24 | 24 | $29x+114x^2$ | 2 | 23/1/1 | 0.1113 | 0.4657 | $55x+342x^2$ | 2 | 27/1/3 | 0.0402 | 0.0680 | 4 | 6.85 |
| 464 | 3 | 3 | 16 | 20 | $247x+58x^2$ | 3 | 28/3/6 | 0.1053 | 0.1920 | $97x+116x^2$ | 3 | 29/1/1 | 0.0130 | 0.0407 | 4 | 4.72 |
| 472 | 3 | 3 | 24 | 24 | $29x+118x^2$ | 2 | 27/2/4 | 0.0523 | 0.1271 | $147x+354x^2$ | 2 | 27/1/3 | 0.0435 | 0.0777 | 4 | 1.64 |
| 480 | 3 | 3 | 30 | 30 | $89x+180x^2$ | 3 | 26/2/4 | 0.1321 | 0.3745 | $209x+120x^2$ | 3 | 27/1/1 | 0.0220 | 0.0919 | 4 | 4.08 |
| 488 | 3 | 3 | 24 | 16 | $91x+122x^2$ | 2 | 27/2/4 | 0.0714 | 0.1743 | $107x+122x^2$ | 2 | 29/1/1 | 0.0108 | 0.0329 | 4 | 5.30 |
| 496 | 3 | 3 | 18 | 10 | $157x+62x^2$ | 3 | 24/1/2 | 0.2355 | 0.4767 | $115x+372x^2$ | 3 | 28/1/4 | 0.0394 | 0.0546 | 4 | 8.73 |
| 504 | 3 | 3 | 28 | 18 | $55x+84x^2$ | 3 | 29/1/1 | 1.5408 | 3.8857 | $55x+168x^2$ | 3 | 30/2/8 | 0.0451 | 0.0565 | 12 | 68.77 |
| 512 | 3 | 3 | 32 | 14 | $31x+64x^2$ | 3 | 27/1/1 | 0.0906 | 0.2295 | $115x+384x^2$ | 3 | 30/1/2 | 0.0225 | 0.0411 | 4 | 5.58 |
| 528 | 3 | 3 | 18 | 22 | $17x+66x^2$ | 3 | 23/1/1 | 0.2698 | 0.8499 | $35x+396x^2$ | 3 | 29/1/3 | 0.0305 | 0.0492 | 4 | 17.27 |
| 544 | 3 | 3 | 32 | 16 | $35x+68x^2$ | 3 | 27/1/1 | 0.0367 | 0.1283 | $41x+408x^2$ | 3 | 32/1/2 | 0.0037 | 0.0105 | 4 | 12.22 |
| 560 | 3 | 3 | 22 | 22 | $227x+420x^2$ | 2 | 29/1/1 | 0.0082 | 0.0307 | $37x+420x^2$ | 2 | 29/1/3 | 0.0157 | 0.0278 | 4 | 1.10 |
| 576 | 3 | 3 | 32 | 18 | $65x+96x^2$ | 3 | 25/1/3 | 0.1678 | 0.3336 | $127x+432x^2$ | 3 | 33/7/29 | 0.0175 | 0.0241 | 2 | 13.84 |
| 592 | 3 | 3 | 20 | 16 | $19x+74x^2$ | 3 | 23/1/1 | 0.1272 | 0.5910 | $163x+444x^2$ | 3 | 31/1/1 | 0.0082 | 0.0227 | 4 | 26.04 |
| 608 | 2.75 | 3 | 32 | 16 | $37x+76x^2$ | 3 | 30/1/2 | 0.0390 | 0.1261 | $41x+456x^2$ | 3 | 32/2/6 | 0.0123 | 0.0309 | 4 | 4.08 |
| 624 | 2.75 | 3 | 22 | 18 | $41x+234x^2$ | 3 | 23/1/1 | 0.3643 | 1.4958 | $47x+156x^2$ | 3 | 31/1/1 | 0.0065 | 0.0248 | 4 | 60.31 |
| 640 | 2.75 | 3 | 32 | 20 | $39x+80x^2$ | 3 | 31/4/8 | 0.0328 | 0.0965 | $141x+160x^2$ | 3 | 33/3/9 | 0.0144 | 0.0310 | 4 | 3.11 |
| 656 | 2.75 | 3 | 22 | 20 | $185x+82x^2$ | 3 | 29/1/3 | 0.0381 | 0.1053 | $49x+292x^2$ | 3 | 32/1/2 | 0.0140 | 0.0398 | 4 | 2.65 |
| 672 | 2.75 | 3 | 32 | 24 | $43x+252x^2$ | 3 | 27/1/1 | 1.1711 | 3.9473 | $79x+168x^2$ | 3 | 29/1/3 | 0.0202 | 0.0467 | 4 | 84.52 |
| 688 | 2.75 | 3 | 24 | 22 | $21x+86x^2$ | 3 | 23/1/1 | 0.1745 | 0.9186 | $31x+172x^2$ | 3 | 31/1/3 | 0.0129 | 0.0272 | 4 | 33.77 |
| 704 | 2.75 | 3 | 22 | 32 | $155x+44x^2$ | 4 | 22/2/4 | 1.1057 | 3.8430 | $219x+528x^2$ | 2 | 33/1/1 | 0.0021 | 0.0111 | 4 | 346.22 |
| 720 | 2.75 | 3 | 36 | 18 | $79x+120x^2$ | 3 | 31/2/6 | 0.0235 | 0.0586 | $19x+180x^2$ | 2 | 31/1/3 | 0.0096 | 0.0293 | 4 | 2.00 |
| 736 | 2.75 | 3 | 32 | 22 | $139x+92x^2$ | 3 | 27/1/1 | 0.0477 | 0.1853 | $55x+552x^2$ | 2 | 33/1/3 | 0.0061 | 0.0153 | 4 | 12.11 |
| 752 | 2.75 | 3 | 26 | 20 | $23x+94x^2$ | 3 | 23/1/1 | 0.1795 | 0.9793 | $207x+564x^2$ | 2 | 33/1/1 | 0.0042 | 0.0153 | 4 | 64.01 |
| 768 | 2.75 | 3 | 24 | 32 | $217x+48x^2$ | 4 | 24/2/4 | 0.4168 | 1.4476 | $47x+672x^2$ | 3 | 31/2/6 | 0.0133 | 0.0353 | 4 | 41.01 |
| 784 | 2.75 | 3 | 26 | 22 | $25x+98x^2$ | 3 | 27/1/1 | 1.1551 | 4.5595 | $59x+196x^2$ | 2 | 33/1/1 | 0.0041 | 0.0132 | 4 | 345.32 |
| 800 | 2.75 | 3 | 32 | 14 | $17x+80x^2$ | 5 | 31/3/9 | 0.0334 | 0.0950 | $187x+600x^2$ | 2 | 32/1/4 | 0.0060 | 0.0166 | 4 | 5.72 |
| 816 | 2.75 | 3 | 26 | 18 | $127x+102x^2$ | 3 | 24/1/2 | 0.1721 | 0.7021 | $95x+612x^2$ | 2 | 32/1/4 | 0.0072 | 0.0166 | 4 | 42.30 |
| 832 | 2.75 | 1 | 26 | 26 | $25x+52x^2$ | 4 | 23/1/1 | 0.0851 | 0.7081 | $181x+624x^2$ | 2 | 33/1/3 | 0.0020 | 0.0055 | 16 | 128.75 |
| 848 | 2.75 | 1 | 28 | 18 | $239x+106x^2$ | 3 | 29/2/6 | 0.0272 | 0.0769 | $179x+636x^2$ | 2 | 35/1/3 | 0.0007 | 0.0021 | 4 | 36.62 |

| | | | | | | | | | | | | | | | | | |
|---|---|---|---|---|---|---|---|---|---|---|---|---|---|---|---|---|---|
| 864 | 2.75 | 1 | 32 | 28 | $17x+48x^2$ | 8 | 28/2/4 | 0.0289 | 0.1248 | $115x+648x^2$ | 2 | 34/1/4 | 0.0016 | 0.0034 | 8 | 36.71 |
| 880 | 2.75 | 1 | 28 | 20 | $137x+110x^2$ | 3 | 25/1/1 | 0.0304 | 0.2675 | $47x+660x^2$ | 2 | 35/1/3 | 0.0007 | 0.0021 | 4 | 127.38 |
| 896 | 2.75 | 1 | 18 | 18 | $215x+112x^2$ | 3 | 31/2/6 | 0.0097 | 0.0290 | $311x+224x^2$ | 2 | 35/1/3 | 0.0007 | 0.0021 | 8 | 13.81 |
| 912 | 2.75 | 1 | 30 | 18 | $29x+114x^2$ | 3 | 29/1/3 | 0.0126 | 0.0383 | $43x+684x^2$ | 2 | 35/1/3 | 0.0007 | 0.0021 | 4 | 18.24 |
| 928 | 2.75 | 1 | 16 | 20 | $15x+58x^2$ | 4 | 28/2/4 | 0.0268 | 0.1244 | $329x+580x^2$ | 3 | 34/1/2 | 0.0007 | 0.0034 | 4 | 36.59 |
| 944 | 2.75 | 1 | 30 | 26 | $147x+118x^2$ | 3 | 27/2/4 | 0.0428 | 0.2020 | $71x+708x^2$ | 2 | 35/1/3 | 0.0007 | 0.0021 | 8 | 96.19 |
| 960 | 2.5 | 1 | 30 | 30 | $29x+60x^2$ | 4 | 26/1/2 | 0.0602 | 0.2887 | $91x+720x^2$ | 2 | 33/1/3 | 0.0035 | 0.0112 | 22 | 25.78 |
| 976 | 2.5 | 1 | 32 | 32 | $59x+122x^2$ | 3 | 30/2/4 | 0.0185 | 0.0902 | $181x+732x^2$ | 2 | 35/1/3 | 0.0014 | 0.0044 | 4 | 20.50 |
| 992 | 2.25 | 1 | 22 | 24 | $65x+124x^2$ | 3 | 22/2/4 | 1.1803 | 5.8544 | $351x+248x^2$ | 2 | 35/1/1 | 0.0009 | 0.0092 | 4 | 636.35 |
| 1008 | 2 | 1 | 28 | 36 | $55x+84x^2$ | 4 | 30/322/644 | 9.9076 | 49.935 | $313x+252x^2$ | 2 | 35/1/3 | 0.0056 | 0.0187 | 4 | 2670.32 |

Legend (for lengths from 40 up to 512):

| | |
|---|---|
| ▨ | QPP with TUB(FER) smaller than that of LTE-QPP and as good as that of LS-QPP-TUB(FER)min |
| 🟪 | QPP with TUB(FER) smaller than those of LTE-QPP and LS-QPP-TUB(FER)min |
| 🟫 | QPP with TUB(FER) close to that of LTE-QPP and smaller than that of LS-QPP-TUB(FER)min |
| 🟨 | QPP with TUB(FER) grater than that of LTE-QPP and equal to that of LS-QPP-TUB(FER)min |

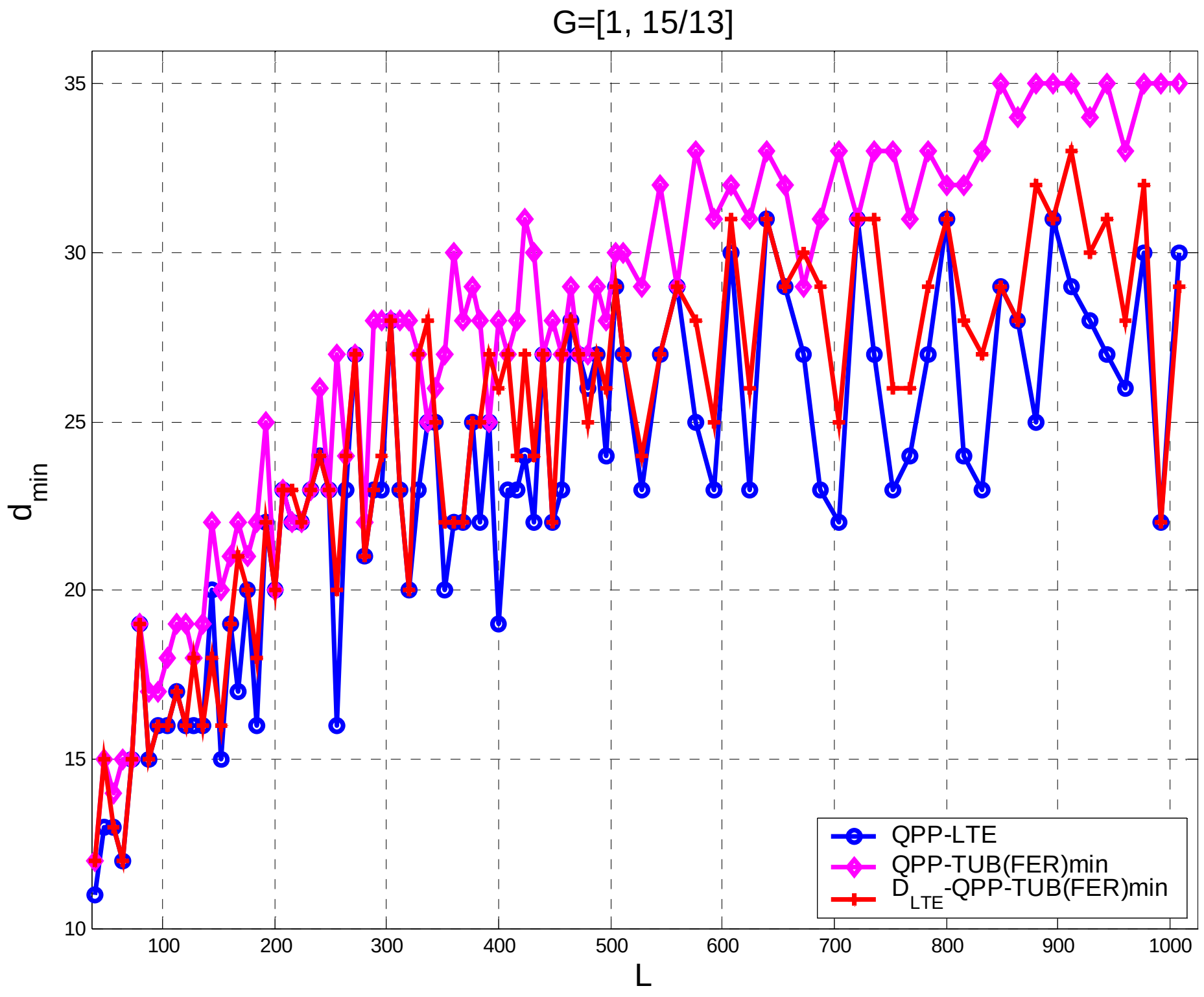

Fig. 1. Minimum distances of turbo codes with LTE –QPP, QPP-TUB(FER)min and $D_{LTE}$-QPP-TUB(FER)min interleavers with lengths from the LTE standard, from 40 up to 1008

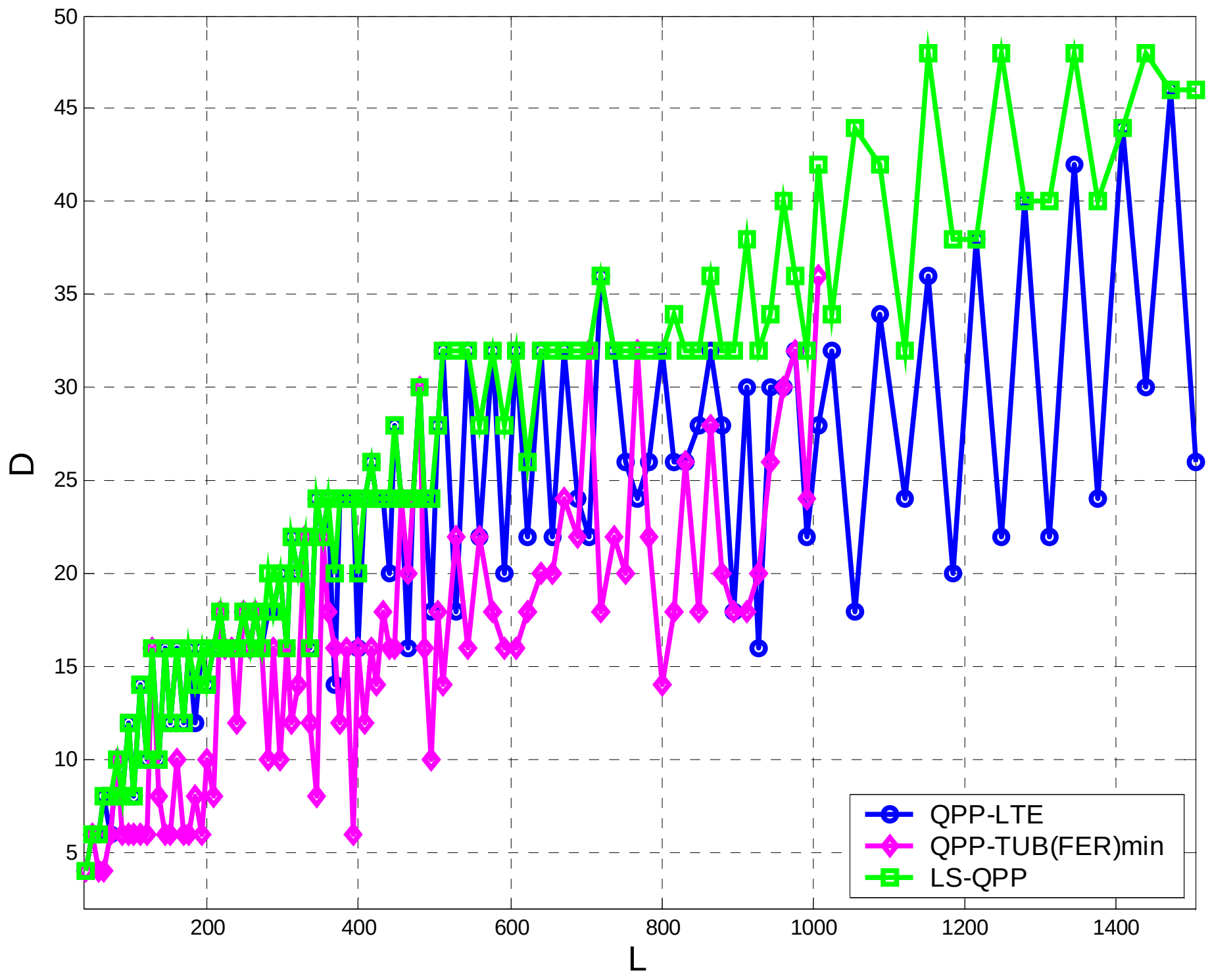

Fig. 2. Parameter D for LTE –QPP and LS-QPP interleavers with lengths from the LTE standard from 40 up to 1504 and for QPP-TUB(FER)min interleavers with lengths from 40 up to 1008

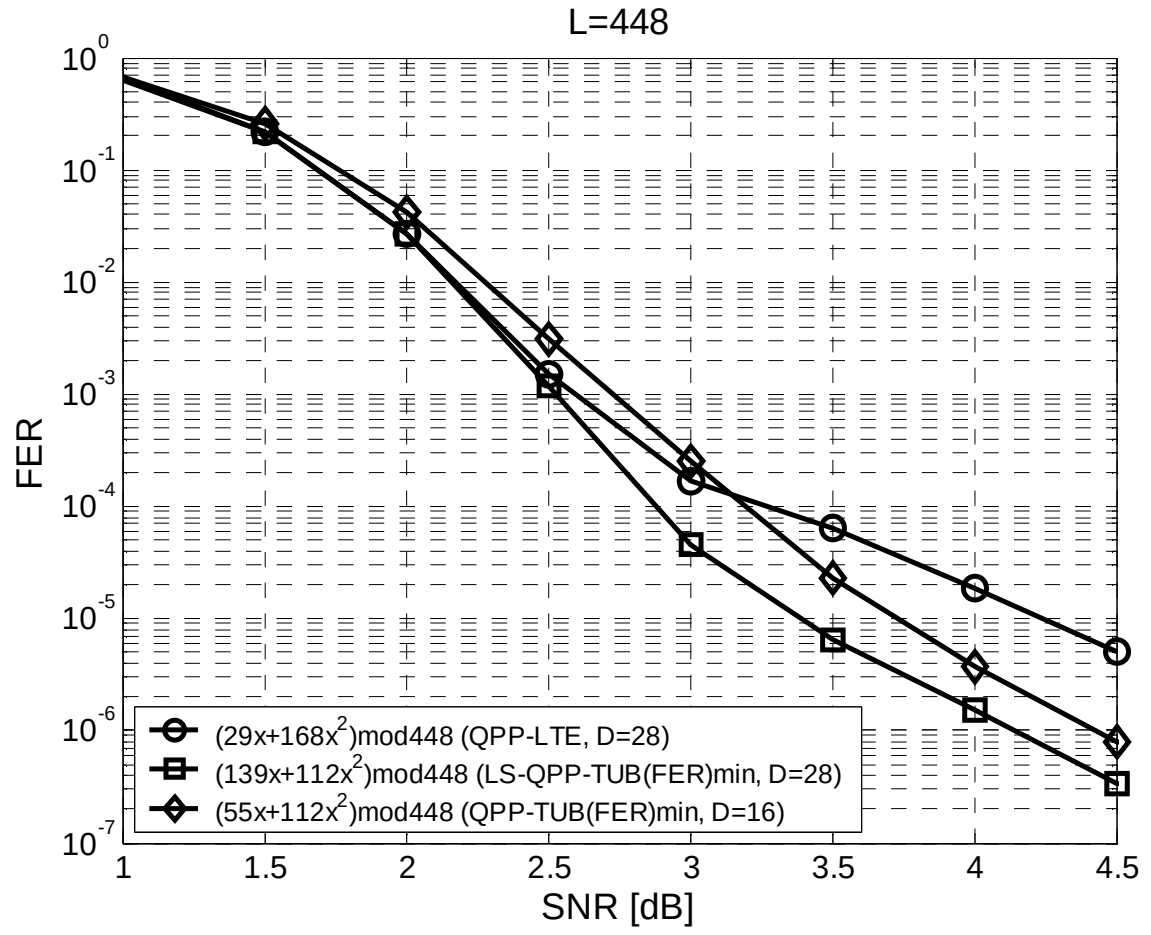

Fig. 3. FER curves for interleavers of lengths 448

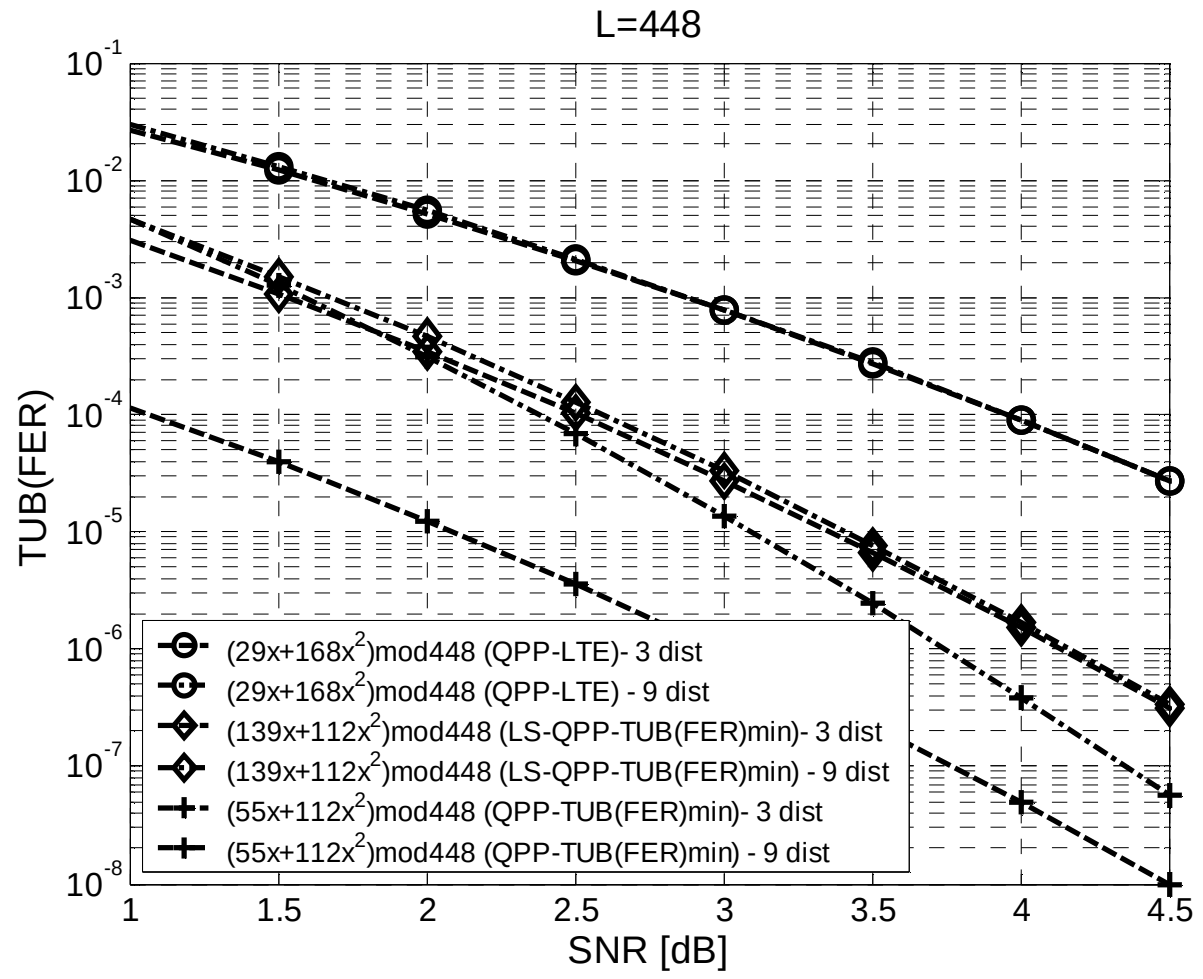

Fig. 4. TUB(FER) curves for interleavers of lengths 448

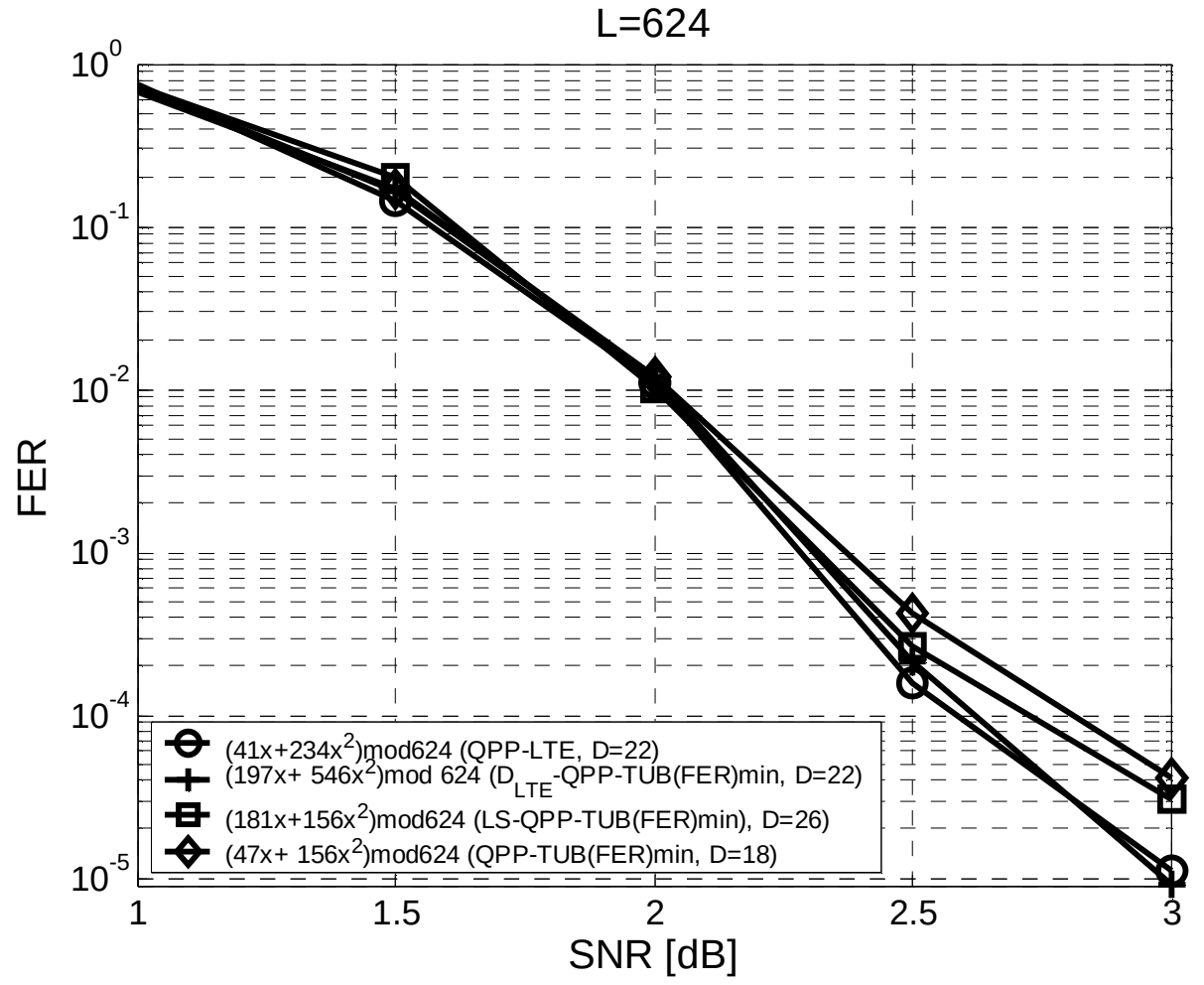

Fig. 5. FER curves for interleavers of lengths 624

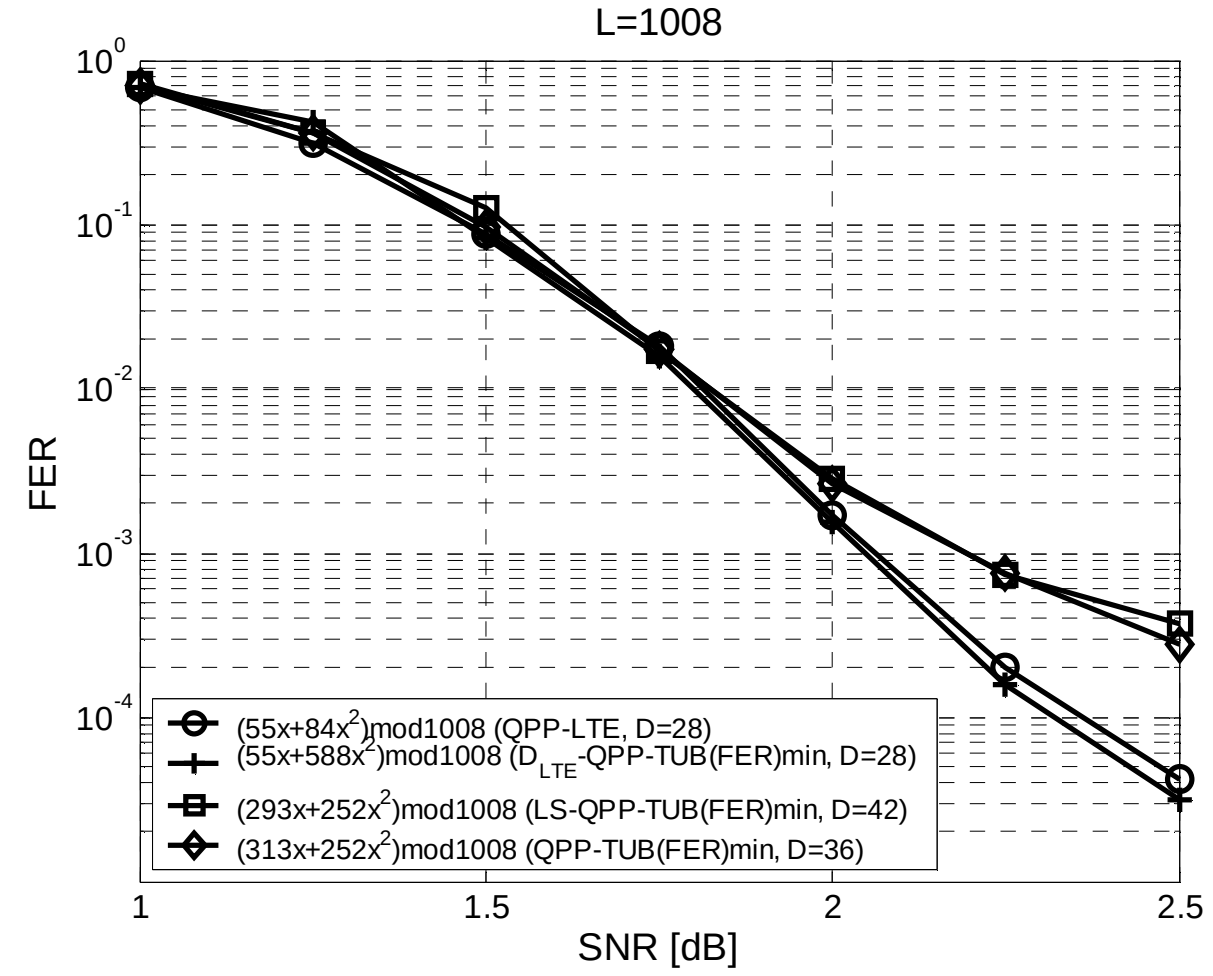

Fig. 6. FER curves for interleavers of lengths1008

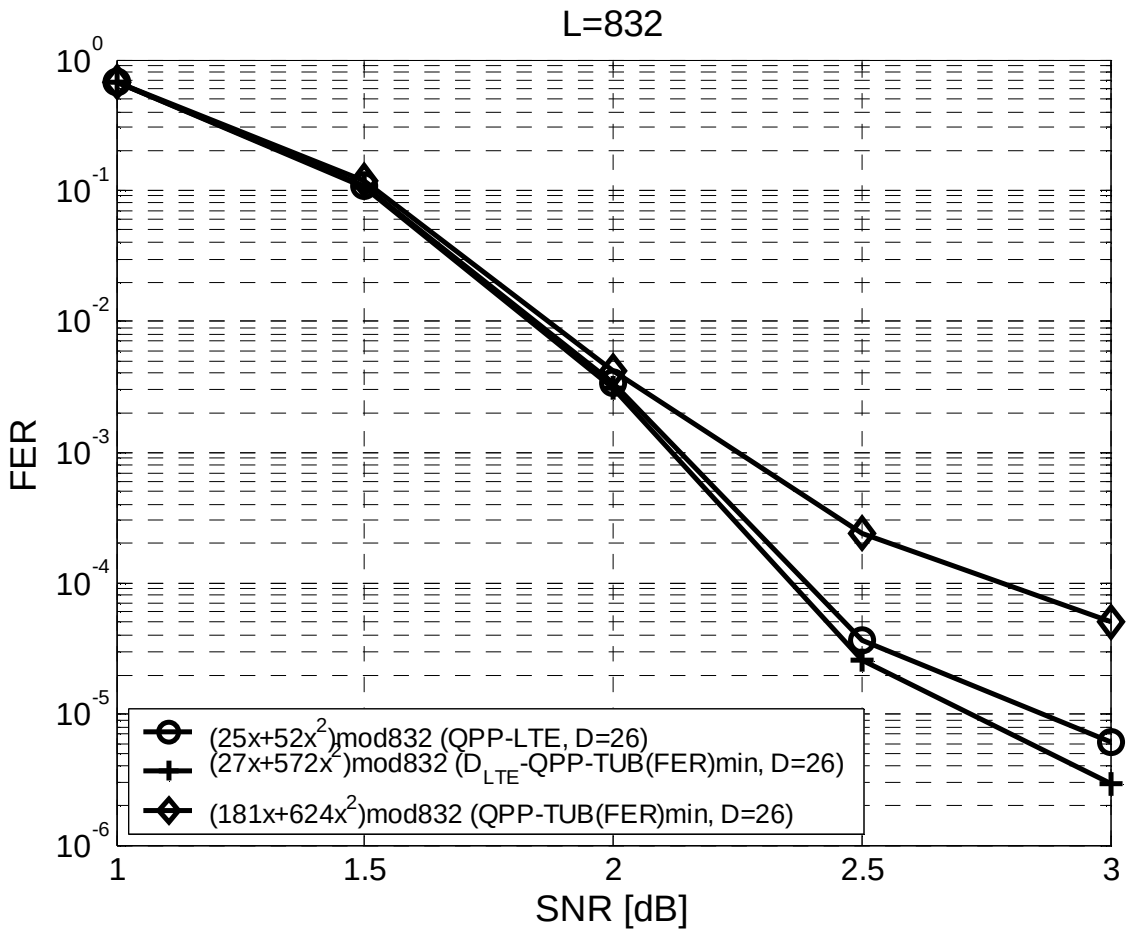

Fig. 7. FER curves for interleavers of lengths 832